\documentclass[twocolumn, float fix, prb, aps, showpacs, longbibliography]{revtex4-2}
\usepackage{graphicx, amssymb, color, amsmath}
\usepackage{nicefrac}
\usepackage{multirow, array, booktabs}
\usepackage{romannum}
\usepackage{mathtools}
\usepackage{bbm}
\usepackage{mathrsfs}
\usepackage{IEEEtrantools}
\usepackage{caption}
\usepackage{subcaption}
\usepackage[font=small,labelfont=bf,
   justification=raggedright,
   format=plain]{caption}
\usepackage[titletoc, title]{appendix}
\usepackage[colorlinks, bookmarks=true, citecolor=blue, linkcolor=red, urlcolor=blue]{hyperref}
\usepackage{orcidlink}
\usepackage{braket}
\usepackage{float}
\usepackage{bm}
\usepackage{tikz}
\usepackage{pgfplots}
\usepackage{pgf}
\usepackage{nicematrix}
\usepackage{diagbox}
\usepackage{lipsum} 
\AtBeginDocument{\pagenumbering{arabic}}

\begin{document}

\title{Numerical demonstration of Abelian fractional statistics of composite fermion excitations in the spherical geometry}

\author{Koyena Bose}
\email{koyenab@imsc.res.in}
\affiliation{Institute of Mathematical Sciences, CIT Campus, Chennai, 600113, India}
\affiliation{Homi Bhabha National Institute, Training School Complex, Anushaktinagar, Mumbai 400094, India}

\author{Ajit C. Balram\orcidlink{0000-0002-8087-6015}}
\email{cb.ajit@gmail.com}
\affiliation{Institute of Mathematical Sciences, CIT Campus, Chennai, 600113, India}
\affiliation{Homi Bhabha National Institute, Training School Complex, Anushaktinagar, Mumbai 400094, India}

\date{\today}

\begin{abstract}

Fractional quantum Hall (FQH) fluids host quasiparticle excitations that carry a fraction of the electronic charge. Moreover, in contrast to bosons and fermions that carry exchange statistics of $0$ and $\pi$ respectively, these quasiparticles of FQH fluids, when braided around one another, can accumulate a Berry phase, which is a fractional multiple of $\pi$. Deploying the spherical geometry, we numerically demonstrate that composite fermion particle (CFP) excitations in the Jain FQH states carry Abelian fractional statistics. Previously, the exchange statistics of CFPs were studied in the disk geometry, where the statistics get obscured due to a shift in the phase arising from the addition of another CFP, making its determination cumbersome without prior knowledge of the shift. We show that on the sphere this technical issue can be circumvented and the statistics of CFPs can be obtained more transparently. The ideas we present can be extended to determine the statistics of quasiparticles arising in certain non-Abelian partonic FQH states.

\end{abstract}	
\maketitle

\section{Introduction}

Topology allows for the possibility that the exchange of identical particles in two dimensions leads to a Berry phase $e^{i\pi\Theta}$, where $\Theta$ takes a fractional value~\cite{Leinaas77}. This concept of Abelian anyonic statistics was only of theoretical interest up until it was realized that anyons materialize in experimentally accessible condensed matter systems which are effectively two-dimensional. One such system is fractional quantum Hall (FQH) fluids~\cite{Tsui82} that exhibit exotic topological properties, foremost among which is hosting anyonic excitations that carry fractional charge~\cite{Laughlin83} and fractional statistics~\cite{Wilczek82, Arovas84, Halperin84}. The fractional charge in an FQH fluid was observed two and a half decades ago~\cite{de-Picciotto97} but an experimental demonstration of fractional statistics remained elusive. Very recently, independent experimental observations deploying different techniques have measured fractional statistics in certain FQH liquids~\cite{Nakamura20, Bartolomei20, Nakamura23, kim2024aharonovbohm, Samuelson24, Werkmeister24} (see also the related perspective of Ref.~\cite{Read23} and work of Ref.~\cite{Kivelson24}). Besides Abelian fractional statistics, a few FQH states are believed to host anyons that carry more enigmatic non-Abelian fractional statistics where their exchange not only accumulates a fractional Berry phase but the state itself changes. These non-Abelian anyons can potentially serve as building blocks for carrying out fault-tolerant topological quantum computation~\cite{Freedman03}.

In the lowest Landau level (LLL), the FQH effect predominantly occurs at filling factors $\nu{=}n/(2pn{\pm}1)$, where $n$ and $p$ are positive integers~\cite{Eisenstein90a}. The appearance of these fractions can be understood using the composite fermion (CF) theory~\cite{Jain89} which postulates that the FQH effect of electrons is the integer quantum Hall (IQH) effect of CFs, which are bound states of electrons and an even number ($2p$) of vortices. These FQH states, known as Jain states, host CF particle (CFP) and CF hole (CFH) excitations on top of the ground state, which are fractionally charged Abelian anyons that when braided around one another lead to the accumulation of fractional Berry phase in the many-body wave function. 

Numerical studies were first carried out to calculate the fractional statistics of the Laughlin quasiholes (identical to CFHs) and quasiparticles (microscopically different from CFPs~\cite{Jain07}) of the 1/3 Laughlin state in the disk geometry~\cite{Laughlin83, Kjonsberg99}. While the quasihole statistics and charge agreed well with the theoretically expected result, the computed statistics for the quasiparticle did not converge to the expected value despite showing an accurate fractional charge. Recently, it has been understood that Laughlin's quasiparticles do not carry the right quantum number (fractional ``spin") to serve as anti-anyons to Laughlin quasiholes~\cite{Nardin23}. Thereafter, these studies were extended to the more general Jain states in the disk geometry where the statistics of the CFPs converged to well-defined values. Although the statistics of the Jain CFP state were found to be correct in magnitude, it had an opposite sign to what was anticipated from theory~\cite{Kjonsberg99b}. Following this, Jeon \emph{et al}.~\cite{Jeon04} showed that a previously unaccounted shift in the position of a CFP due to the addition of other CFPs was responsible for obscuring the statistics. They resolved this issue by providing a relation for the shift which was reconfirmed from the computation of the one-particle density at 1/3. Additionally, the braiding statistics computed for CFPs in disk geometry suffer from edge effects which require going to large systems so that statistics are evaluated for CFPs that are deep inside the bulk.

Compact geometries, such as a sphere or torus, that are devoid of an edge are ideally suited to compute braiding statistics which is a bulk property. The torus geometry results in a degeneracy of FQH states stemming from its non-trivial genus, while states on the sphere are easier to work with since they are non-degenerate. In this work, we compute statistics of CFPs in the spherical geometry by circulating one around another along latitudes as shown in Fig.~\ref{fig: schematic_CFP_phase}. This numerical evaluation is carried out using the method developed by Tserkovnyak and Simon~\cite{Tserkovnyak03}. Unlike in the disk geometry, it turns out that statistics can be extracted on the spherical geometry without the need to know the exact form of the shift a priori. Using the spherical geometry, we evaluate the statistics of CFPs for the 1/3, 1/5, 2/5, 2/9, and 3/7 Jain states and find that these are in agreement with theoretical predictions. On the disk, the analogous computation for 1/5, 2/9, and 3/7 was not done previously presumably because the sizes of the CFPs in these states are larger than those at 1/3 and 2/5~\cite{Jain07, Balram13b} and require going to very large systems.

The article is organized as follows: In Sec.~\ref{sec: cf theory}, we introduce the CF theory, paying special attention to its formulation in the spherical geometry. We then discuss the Berry phase accumulated by CFs upon circumscribing a loop. This requires the construction of coherent states for representing localized particles in real space and we provide a primer on that. In Sec.~\ref{sec: operations}, we define various quasiparticle operations on the sphere and provide theoretical expectations for the Berry phase accumulated corresponding to these operations. Thereafter, we discuss a numerical technique that can be used to extract the statistics from the Berry phase. In Sec.~\ref{sec: numerical}, we present numerical results for the looping operations and the extracted statistics. We conclude the paper in Sec.~\ref{sec: discussion} with a discussion of the results and an outlook for the future.

\section{Composite fermion theory}
\label{sec: cf theory}
The central postulate of the composite fermion (CF) theory~\cite{Jain89} is that strongly interacting electrons in a topologically-ordered flat band such as a Landau level transform into new emergent particles, \textit{composite fermions}, which are themselves nearly noninteracting. Composite fermions are formed by attaching an even number of vortices to electrons. Due to the vortex attachment, CFs experience an effective magnetic field $B^*$ that is smaller compared to the external magnetic field $B$. The CF theory maps an FQH problem of electrons at $\nu{=}n{/}(2pn {\pm} 1)$ in a magnetic field $B$ to an integer quantum Hall (IQH) state of CFs filling $n$ CF-Landau-like levels [termed $\Lambda$ levels ($\Lambda$Ls)] in the effective field $B^{*}$.

Using the CF theory, the Jain wave function for the ground state of electrons at $\nu{=}n/(2pn{\pm}1)$ is written as~\cite{Jain89}
\begin{equation}
    \Psi_{\nu}= \mathcal{P}_{\rm LLL} \Phi_{\pm n} \Phi_1^{2p},
    \label{eq: cf states}
\end{equation}
where $\Phi_n$ is the Slater determinant wave function of $n$ filled LLs of electrons ($\Phi_{{-}n}{\equiv}\Phi_{\bar{n}}{=}[\Phi_{n}]^{*}$) and the Laughlin-Jastrow factor $\Phi^{2p}_1{\equiv}\prod_{i<j}(z_{i}-z_{j})^{2p}$ [$z_{j}{=}x_{j}{-}iy_{j}$ is the two-dimensional coordinate of the $j^{\rm th}$ electron parametrized as a complex number] binds $2p$ vortices to each electron to turn them into CFs. Appropriate to the high magnetic field of our interest, the product is then projected to the lowest LL (LLL) which the operator $\mathcal{P}_{\rm LLL}$ accomplishes. Furthermore, excitations over the ground state are produced either by creating CFPs, i.e., placing CFs in the lowest empty $\Lambda$L, or CFHs, i.e., by removing CFs from the topmost filled $\Lambda$L. Their wave functions are given by

\begin{equation}
    \Psi^{N_{\rm CFPs}}_{\nu}= \mathcal{P}_{\rm LLL} \Phi^{N_{\rm CFPs}}_{\pm n} \Phi_1^{2p},
    \label{eq: cf QP states}
\end{equation}

\begin{equation}
    \Psi^{N_{\rm CFHs}}_{\nu}= \mathcal{P}_{\rm LLL} \Phi^{N_{\rm CFHs}}_{\pm n} \Phi_1^{2p},
    \label{eq: cf QH states}
\end{equation}
where $N_{\rm CFPs}$ and $N_{\rm CFHs}$ refer to the number of CFPs and CFHs present in the state respectively. Although projected and unprojected states are microscopically different, studies show that they describe the same underlying topological phase~\cite{Balram16b, Anand22}. Unless otherwise stated, in this work, we will be using the more easily computable unprojected wave functions to evaluate the Berry phase accumulated from different operations. Next, we provide a background on spherical geometry~\cite{Haldane83} which we will use throughout this work.

\subsection{Spherical geometry}
A two-dimensional system of $N$ electrons subjected to a perpendicular magnetic field can be simulated by placing the electrons on the surface of a sphere in the presence of a radial magnetic flux of strength $2Q\phi_0$ ($2Q$ is an integer), $\phi_0{=}hc{/}e$ is the magnetic flux quantum, generated by a magnetic monopole sitting at the center of the sphere. This sphere, known as the Haldane sphere~\cite{Haldane83}, has a radius $R{=}\sqrt{Q}\ell$, where $\ell{=}\sqrt{\hbar c{/}(eB)}$ is the magnetic length at field $B{=}2Q\phi_0{/}(4\pi R^2)$. We choose the vector potential, $\mathbf{A}{=}(-\hslash c Q{/}eR)\cot \theta \hat{\bm{\phi}}$, wherein the monopole can be viewed as two Dirac strings of strength $Q\phi_{0}$ passing through the north and south poles. With this choice of gauge, a position on the sphere $\Omega{=}(u,v)$ is specified by the spinor coordinates $u{=}\cos({\theta{/}2})e^{i\phi{/}2}$ and $v{=}\sin({\theta{/}2})e^{{-}i\phi{/}2}$, where $\theta$ and $\phi$ are the polar and azimuthal angles on the sphere. Due to the curvature in the spherical geometry, incompressible states such as the IQH and FQH fluids occur on it when $2Q{=}\nu^{{-}1}N{-}\mathcal{S}$, where the Wen-Zee shift $\mathcal{S}$~\cite{Wen92} is a topological quantity that characterizes the fluid. 

On the sphere, an IQH state with $n$ filled LLs occurs at $2Q_n{=}(N{/}n{-}n)$ with $N$ divisible by $n$ and $N{\geq}n^2$. Therefore, the FQH state at $\nu{=}n{/}(2pn{\pm}1)$ with $n$ filled $\Lambda$Ls of CFs [see Eq.~\eqref{eq: cf states}] occurs at 
\begin{equation}
    2Q={\pm}2Q_n+2p2Q_1.
    \label{eq: fqhe_gs_flux}
\end{equation}
Here, the flux of $2Q_{1}$ results from the vortex attachment done by the Jastrow factor, which on the sphere takes the form $\Phi_{1}{=}\prod_{i<j}(u_{i} v_{j} {-} u_{j} v_{i})$, where $u_{j}$ and $v_{j}$ are the spinor coordinates of the $j^{\rm th}$ electron. Under the addition of CFPs, the effective flux has to be adjusted as $2Q_n^{N_{\rm CFPs}}{=}(N{/}n{-}N_{\rm CFPs}{/}n{-}n)$ which results in a total flux 
\begin{equation}
    2Q^{N_{\rm CFPs}}={\pm}2Q_n^{N_{\rm CFPs}}+2p2Q_1.
    \label{eq: fqhe flux}
\end{equation}
From here on in, unless otherwise stated, we shall restrict ourselves to the $n/(2pn{+}1)$ Jain states.

\subsection{Berry phase of composite fermions}
When a CF goes around in a closed loop enclosing an area $A$, the Berry phase that the many-body wave function accumulates is $e^{i\xi}$, where
\begin{equation}
    \xi =-2\pi \frac{BA}{\phi_0}+2\pi~2p N_{\rm enc}.
    \label{eq: fqhe_Berry_phase}
\end{equation}
The first term is the Aharonov-Bohm (AB) phase accumulated when an electron encircles a loop of area $A$ in the presence of a magnetic field $B$. The second term arises from the $2p$ vortices seen by the looping electron on each of the other $N_{\rm enc}$ electrons within the enclosed area, with each of them contributing a phase of $2\pi$. This accumulated phase can be viewed as an AB phase for the CFs
\begin{equation}
    \xi^{*} =-\frac{2\pi B^{*} A}{\phi_0}
    \label{eq: fqhe_eff_Berry_phase}
\end{equation}
experiencing an effective magnetic field $B^{*}$ given by 
\begin{equation}
   B^{*} =B- 2p\rho\phi_0,
    \label{eq: eff-mag field}
\end{equation}
where $\rho{=}N/(4\pi Q\ell^{2})$ [in the thermodynamic limit $\rho{=}\nu/(2\pi \ell^2)$] is the average density of the system. 

\begin{figure}[t!]
\begin{center}
\includegraphics[width=0.5\textwidth,height=0.28\textwidth]{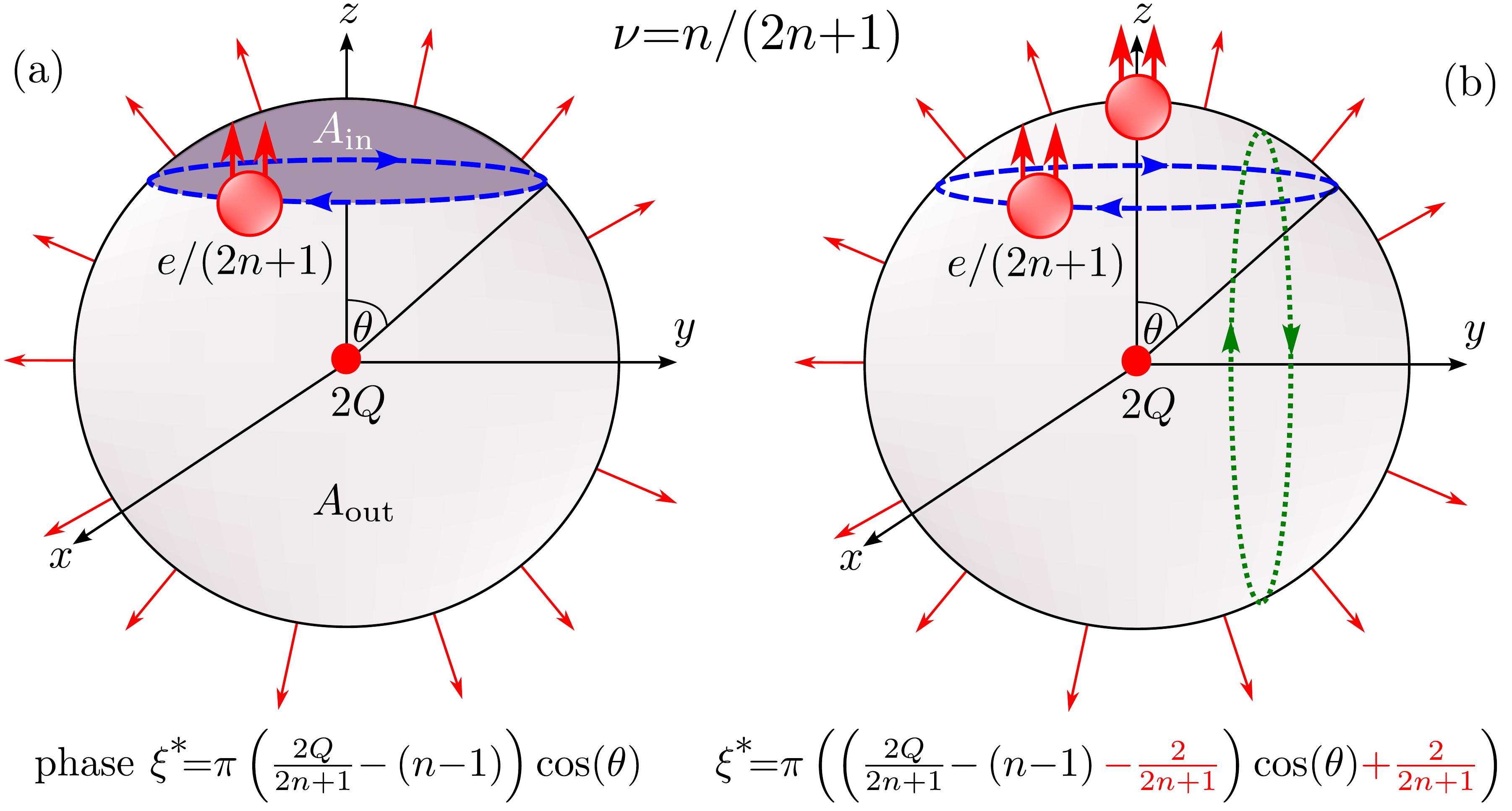}
\caption{Pictorial depiction of the phase, $\xi^{*}$, accumulated when a CF particle (electron and two vortices depicted by a ball and two vertical arrows) loops on the Haldane sphere. Red arrows show flux emanating from the magnetic monopole of strength $2Q$ placed at the origin that coincides with the sphere's center. (a) A single CF particle moving along the latitude $\theta$ picks up an Aharonov-Bohm phase proportional to the area of the loop it encloses. (b) A CF particle moving along latitude $\theta$ in the presence of another CF particle static at the north pole picks up an additional phase (shown in red). The green dotted loop shows an example of a path wherein the singularities of the vector potential at the north and south poles are placed on the same side of the path.}
\label{fig: schematic_CFP_phase}
\end{center}
\end{figure}

This relation of the Berry phase can be rewritten in terms of the flux but one has to do this carefully on the sphere since a mere replacement of phase with flux does not give the correct result due to the singularities in the vector potential. In the spherical geometry, the phase accumulated depends on the trajectory followed by the particle. If the CFP moves along a closed path that places the two singularities at the north and south poles (in the gauge we have chosen) on two different sides of the loop [as in the blue dashed loop in Fig.~\ref{fig: schematic_CFP_phase}(a)], the phase accumulated is 
 \begin{equation}
    \xi^* =-\frac{ Q_n (A_{\rm in}-A_{\rm out})}{2 R^2},
    \label{eq: fqhe_eff_Berry_phase in_out}
\end{equation}
where $A_{\rm in}$ is the area covered within the loop, and $A_{\rm out}$ represents the area outside the loop as shown in Fig.~\ref{fig: schematic_CFP_phase}. However, if the CFP follows a loop that places the singularities on the same side of the loop [as in the green dotted loop in Fig.~\ref{fig: schematic_CFP_phase}(b)], the accumulated phase is
 \begin{equation}
    \xi^* =-\frac{ Q_n A_{\rm in}}{R^2},
    \label{eq: fqhe_eff_Berry_phase in}
\end{equation}
with $Q_n {=} Q{/}(2pn{+}1)$ in the thermodynamic limit. The result given in Eq.~\eqref{eq: fqhe_eff_Berry_phase in} is identical to what one finds in the disk geometry~\cite{Jeon04}.
 
\subsection{Coherent states}
In the spherical geometry, the orbital angular momentum $l$ and its $z$-component $m$ are good quantum numbers. Therefore, the single-particle eigenstate of $m$th orbital in $n$th LL with index $n{-}1$ (for $Q{>}0$) is given by~\cite{Wu76, Wu77, Jain07}
 \begin{eqnarray}
           Y_{Q,l,m}&=& N_{Q,l,m}(-1)^{l-m}v^{Q-m}u^{Q+m}\sum_{s=0}^{l-m} (-1)^s \nonumber \\
    &&  \binom {l-Q} s \binom {l+Q} {l-m-s} (v^*v)^{l-Q-s}(u^*u)^s,
    \label{eq: single-particle states}
 \end{eqnarray}
where $N_{Q,l,m}$ is a normalization constant, $l{=}Q{+}n{-}1$ and $m$ goes from ${-}l$ to ${+}l$. The value of the binomial coefficient $\binom j k$ is set to zero if $k{<}0$ or $k{>}j$. To compute the Berry phase of CFHs or CFPs, we will need to create localized states in real space at $\tilde{\Omega}{=}(\tilde{u},\tilde{v})$ on the sphere in different LLs. The coherent state, which is the maximally localized state that one can make, at $\tilde{\Omega}$ in the LLL ($n{=}0$) is~\cite{Jain07}
 \begin{equation}
    \psi^{0}_{\Omega,\tilde{\Omega}}=\sum_{m=-l}^{+l} \bar{Y}_{QQm}(\tilde{\Omega}) Y_{Q,Q,m}(\Omega)=(\tilde{u}^*u+\tilde{v}^*v)^{2Q}.
    \label{eq: coherent state LLL}
\end{equation}
The coherent state for higher LLs can be obtained by a repeated application of $S^{-}$ (analogous to how $a^{\dagger}$ lowers the value of the $L_{z}$ quantum number on the plane, $S^{-}$ operator lowers the $S^{z}$ eigenvalue on the sphere), the LL raising operator on the LLL coherent state~\cite{Greiter11}. The $S^{-}$ operator is defined as 
 \begin{eqnarray}
      S^{-}= v^* \frac{\partial}{\partial u}-u^*\frac{\partial}{\partial v}
    \label{eq: S_minus}
 \end{eqnarray}
The final result for the coherent state for the LL indexed by $n$, aside from a normalization factor, is 
 \begin{eqnarray}
       \psi^{n}_{\Omega,\tilde{\Omega}} &=&(S^-)^{n}\psi^{0}_{\Omega, \tilde{\Omega}} \nonumber \\
       &=&(\tilde{u}^*u+\tilde{v}^*v)^{2Q-n}(v^*\tilde{u}^*-u^*\tilde{v}^*)^{n}.
    \label{eq: coherent state n LL}
 \end{eqnarray}

\section{Formalism of operations}
\label{sec: operations}

The CFP and CFH excitations of the Jain states carry a local fractional charge of ${\pm}e/(2pn{\pm }1)$, compared to the uniform background. The fractional charge endows the CFPs and CFHs with fractional braiding statistics. Unlike point-like objects, the CFPs and CFHs have a finite extent with a density distribution that oscillates in real space~\cite{Jain07, Balram13, Johri14, Ji24}. Thus, when evaluating their Berry phase, it is imperative to keep them sufficiently apart from one another to avoid any overlap. In this section, we define different operations where a CFP loops around a latitude of the sphere in the absence and presence of another CFP. We also give analytic expressions for the accumulated phase that can be used to extract the statistics provided the excitations are kept sufficiently apart.

In contrast to the Laughlin quasihole(s) state~\cite{Laughlin83}, whose wave function is an explicit function of the coordinate(s) of the quasihole(s), excitations in Jain FQH states are described in terms of additional filled or empty orbitals. Consequently, executing directed operations such as moving a CFP on a loop on the spherical surface necessitates the explicit inclusion of the CFP's coordinates in the many-body wave function. This is accomplished for CFPs by populating the lowest available $\Lambda$L with coherent states. However, since CFHs are vacant states, the explicit inclusion of their coordinates in the wave function is not so straightforward. Hence, in this study, we will focus on CFPs only. Nevertheless, we note that it is possible to extract the statistics of the CFHs by creating a vortex at $(U, V)$ i.e., using the wave function $\Psi^{U, V}_{\nu}{=}\prod_{i} \left(u_i V {-} v_j U \right) \Psi_{\nu}$, where $\Psi_{\nu}$ is given in Eq.~\eqref{eq: cf states}. The vortex effectively splits into $n$ quasiholes which carry identical statistics parameters as the $n$ CFHs, one in each $\Lambda$L~\cite{Balram20}.

\subsection{Single loop operation along latitudes}
\label{subsec: Berry_phase_looping}

At $\nu{=}n/(2pn{+}1)$, when a CFP loops once along a latitude making an angle $\theta$ with the $z$-axis [as shown in Fig.~\ref{fig: schematic_CFP_phase}(a)], following Eq.~\eqref{eq: fqhe_eff_Berry_phase in_out}, we expect an AB phase of
 \begin{equation}
    \xi^* =  \pi  \left(2Q^{N_{\rm CFPs}=1}_n -(n-1) \right) \cos(\theta)
    \label{eq: single qp phase}
\end{equation}
given that it resides in the $(n{+}1)^{\rm th}$ $\Lambda$L (indexed by $n$) of the $\Phi_{n}$ [see Eq.~\eqref{eq: cf QP states}]. The term with the power $2Q{-}n$ in the coherent state given in Eq.~\eqref{eq: coherent state n LL} contains the information of the effective flux seen by the CFP and results in a contribution of $\pi (2Q{-}n)\cos(\theta)$ to the accumulated phase as shown in Eq.~\eqref{eq: single qp phase}. Surprisingly, there is an additional $\mathcal{O}(1)$ contribution to the phase of magnitude $\pi\cos(\theta)$ that we identified by numerically computing the phase for IQH states. The effective flux seen by the CFP is $2Q_n^{N_{\rm CFPs}{=}1}{=}(N{/}n{-}1{/}n{-}n)$.

Next, we place a CFP at the north pole which is kept static while another CFP follows the same trajectory as before moving along a latitude at an angle $\theta$ with the $z$-axis [as shown in Fig.~\ref{fig: schematic_CFP_phase}(b)]. The phase accumulated by the moving CFP in this case is
\begin{equation}
    \xi^* =  \pi \left( \left( 2Q_n^{{N_{\rm CFPs}=2}}-(n-1)-\Theta \right) \cos(\theta) +\Theta \right),
    \label{eq: doub qp phase}
\end{equation}
where $\Theta{=}2p{/}(2pn{+}1)$ is the additional statistics [compared to Eq.~\eqref{eq: single qp phase}] accrued due to the presence of the CFP placed at the north pole. The denominator in the factor $\Theta$ comes from the fractional charge associated with the CFP, while the numerator of $2p$ comes from the vortices. The flux equivalent to $\Theta$ is associated with the static CFP while the remaining flux of $\left(2Q_n^{{N_{\rm CFPs}=2}}{-}(n-1){-}\Theta\right)$ is then evenly distributed over the spherical surface. The CFP that loops along the latitude of $\theta$ accumulates an AB phase from this remaining flux [which is what results in the $\pi \left( 2Q_n^{{N_{\rm CFPs}{=}2}}{-}(n-1){-}\Theta \right) \cos(\theta)$ term in Eq.~\eqref{eq: doub qp phase}]. 

\begin{figure}[t!]
\begin{center}
\includegraphics[width=0.47\textwidth,height=0.3\textwidth]{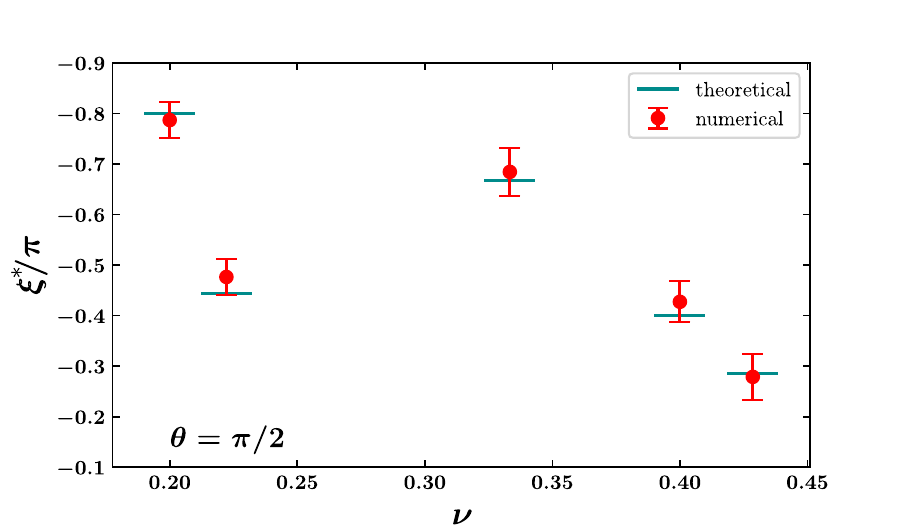}
\caption{Berry phase accumulated when a CFP moves along the equator ($\theta{=}\pi/2$) with another CFP kept static at the north pole for Jain states at $\nu{=}n/(2pn{+}1)$. The expected phase (green line) is equal to ${-}\Theta{=}{-}2p{/}(2pn{+}1)$ and is plotted alongside the numerical result (red dot with error bar determined from the statistical uncertainty of the Monte Carlo simulation) for the largest system size corresponding to each fraction as listed in Table~\ref{tab: frac_stat}.}

\label{fig: frac_stat_equator}
\end{center}
\end{figure}

\subsection{Extraction of statistics}
\label{subsec: extraction_of_statistics}

Surprisingly, it turns out that the phase accumulated by the moving CFP is not as anticipated from Eq.~\eqref{eq: doub qp phase}. This is because a correction must be applied when dealing with multiple CFPs. It was previously shown that when a CFP is added in the presence of another CFP, there is a shift in their positions~\cite{Jeon04}. While the shift in the position of the static CFP is not significant (as long as it stays within $A_{\rm in}$ and far away from the moving CFP), the shift in the position of the moving CFP introduces an additional phase, denoted by $\Delta$, to the overall result. Thus, the net phase accumulated by the moving CFP is 

\begin{equation}
    \frac{\xi^*}{ \pi} =  \left( \left(2Q_n^{N_{\rm CFPs}=2}-(n-1)-\Theta \right) \cos(\theta) +\Theta+ \Delta \right).
    \label{eq: mod_doub_qp_phase}
\end{equation}

 In the disk geometry, the phase accumulated by a CFP moving along a circle of radius $r$ with another CFP kept static at the origin is~\cite{Jeon04}
 \begin{equation}
 \xi^{*,{\rm disk}} {=} 2\pi[{-}r^2{/}2{\ell^{*}}^2{+}\Theta{+}\delta], 
 \label{eq: phase_disk}
 \end{equation}
 where $\delta$ is the additional phase due to the shift in the position of the moving CFP and the effective magnetic length $\ell^{*}{=}(\sqrt{2pn{+}1})\ell$ for filling $\nu{=}n/(2pn{+}1)$~\cite{Jain07}. In the limit $N {\to} \infty$, this phase relation is identical to the one defined in Eq.~\eqref{eq: mod_doub_qp_phase} for the spherical geometry. From the phase relation in the disk geometry, it is clear that accurate knowledge of the extra phase $\delta$ is crucial for extracting statistics. Jeon \emph{et al.}~\cite{Jeon04} found that $\delta{=}{-}4p{/}(2pn{\pm}1){=}{-}2\Theta$ and confirmed this by examining the shift in the position of the CFP by evaluating the one-particle density at 1/3 filling. As $\delta$ is an additional shift in the effective flux and is an intrinsic property of the FQHE state, it is geometry-independent (however, the shift in the position of the moving CFP in different geometries will be different), and should equal the extra phase seen in the spherical geometry ($\Delta{=}\delta$). This equivalence can be further demonstrated by evaluating the phase at $\cos(\theta){=}0$ [equator], where the phase equation simplifies to $\xi^{*}_{\theta{=}\pi/2}/\pi {=} (\Theta{+} \Delta )$ which is consistent with ${-}\Theta$ as depicted in Fig.~\ref{fig: frac_stat_equator}. Utilizing this relation, the phase equation given in Eq.~\eqref{eq: mod_doub_qp_phase} takes the form
\begin{equation}
   \frac{\xi^*}{ \pi}=\left( \left (2Q_n^{N_{\rm CFPs}=2}-(n-1)-\Theta \right ) \cos(\theta) -\Theta \right ).
    \label{eq: mod_doub_qp_phase_final}
\end{equation}
Since the shift is neither position nor geometry-dependent it arises entirely from the flux deletion required to add the extra static CFP and thereby solely depends on the filling. Therefore, even in the absence of the knowledge of the specific value of $\Delta$, as long as we know $\Delta$ is a constant at a given filling (for example, even if we did not know that $\Delta{=}{-}2\Theta$ here), we can find $\Theta$ by subtracting the phase at latitudes $\theta$ and $\pi{-}\theta$ to get
\begin{equation}
   \frac{(\xi_{\theta}^*-\xi_{\pi-\theta}^*)}{ \pi}=2\left( \left(2Q_n^{N_{\rm CFPs}=2}-(n-1)-\Theta \right) \cos(\theta) \right).
    \label{eq: mod_doub_qp_phase_without_delta}
\end{equation}
Importantly, the statistics parameter can be extracted from the $\cos(\theta)$ term, arising due to the curvature of the sphere which protects it from the shift. However, the analogous process on the flat disk is infeasible since $\Theta$ and $\delta$ would have been lost in the subtraction process [see Eq.~\eqref{eq: phase_disk}].

\begin{figure}[htbp!]

\subfloat{}{%
  \includegraphics[clip,width=0.78\columnwidth]{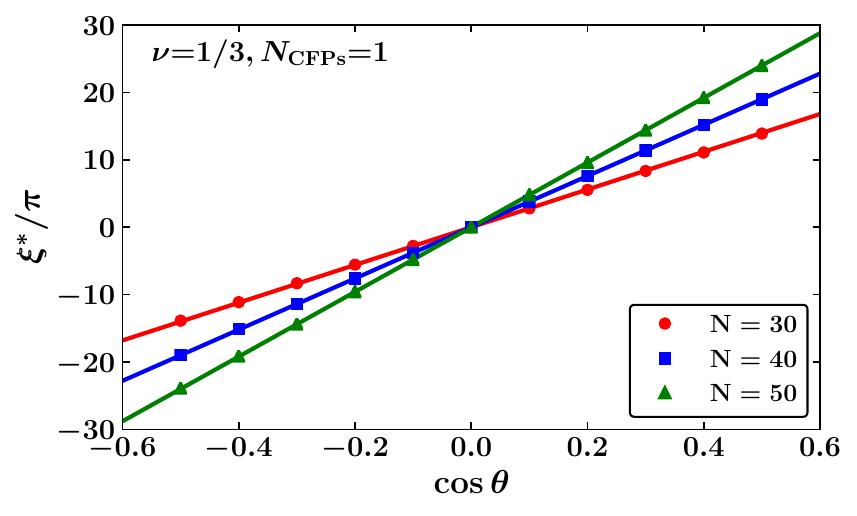}%
   
}

\subfloat{}{%
  \includegraphics[clip,width=0.78\columnwidth]{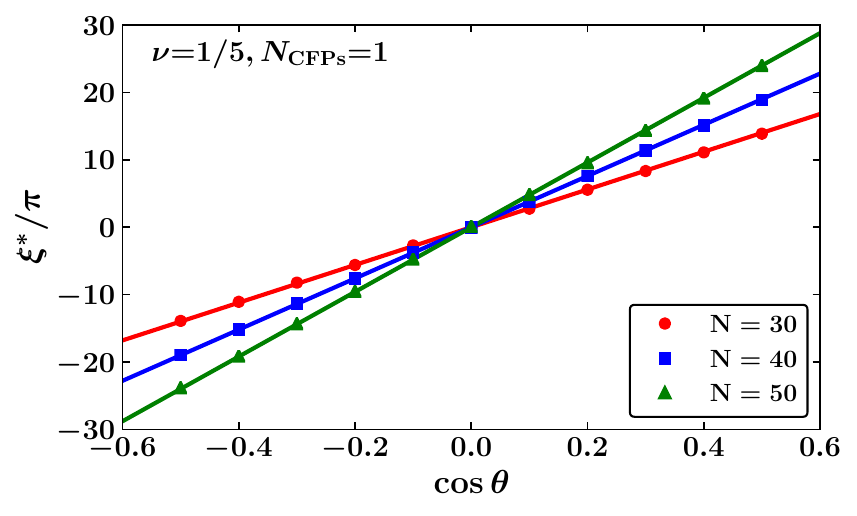}%
  
}

\subfloat{}{%
  \includegraphics[clip,width=0.78\columnwidth]{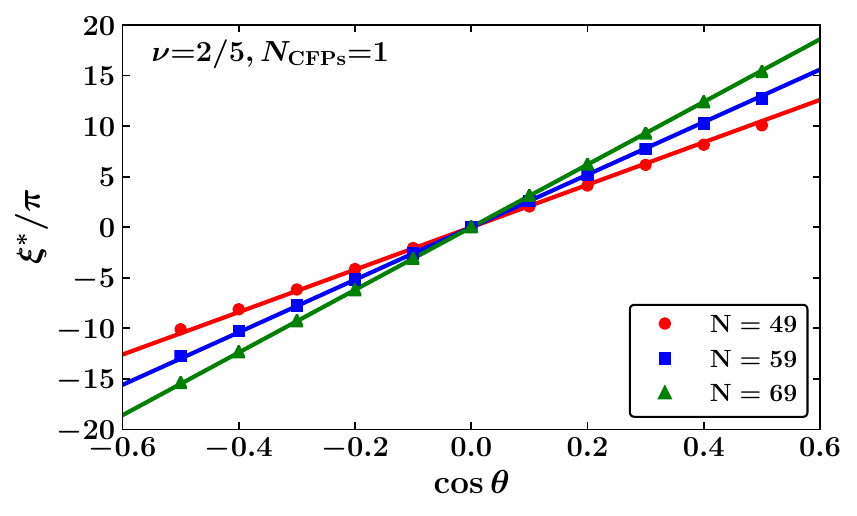}%
   
}

\subfloat{}{%
  \includegraphics[clip,width=0.78\columnwidth]{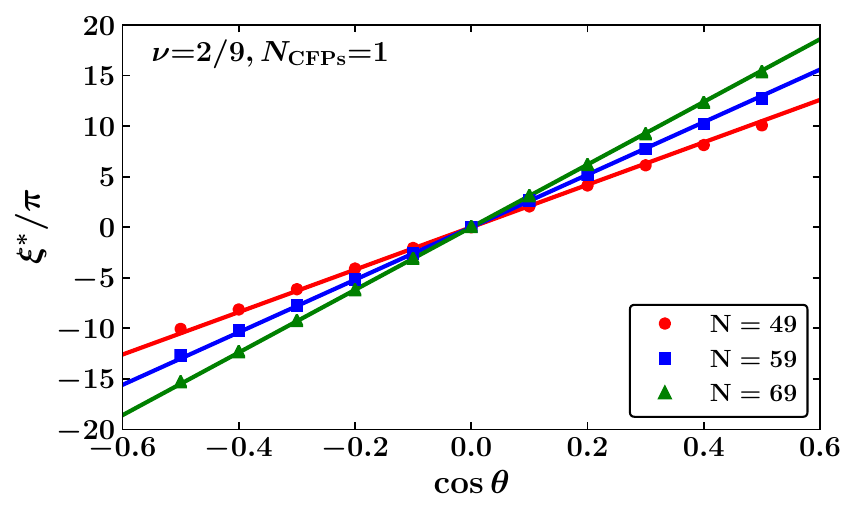}%
  
}

\subfloat{}{%
  \includegraphics[clip,width=0.78\columnwidth]{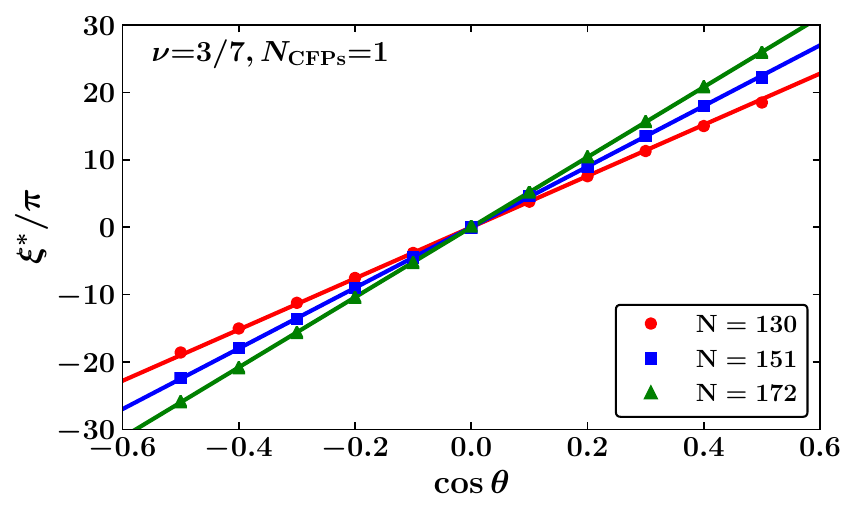}%
  
}

\caption{Berry phase accumulated by a single CFP looping along various latitudes making an angle $\theta$ with the $z$-axis for many filling fractions along the Jain sequence for three different system sizes. Dots and lines correspond to numerically computed phase and theoretically-predicted phases [see Eq.~\eqref{eq: single qp phase}] respectively. }
\label{fig: phase_qp_1}
\end{figure}

\subsection{Operations on many-body wave functions}
\label{subsec: wf_operation}

To compute the statistics of CFPs, it is necessary to carry out relevant operations on the many-body wave function. Accordingly, a Jain state with $N_{\rm CFPs}$ CFPs at filling $\nu{=}n{/}(2pn{+}1)$ can be constructed as 
 \begin{eqnarray}
	\Psi_{\nu}^{N_{\rm CFPs}}&=&\begin{vmatrix}
	\psi^{n}({\Omega_1,\tilde{\Omega}_{N_{\rm CFPs}}}) & \psi^{n}({\Omega_2,\tilde{\Omega}_{N_{\rm CFPs}}}) & . & . & .  \\  

    \psi^{n}({\Omega_1,\tilde{\Omega}_{N_{\rm CFPs}-1}}) & \psi^{n}({\Omega_2,\tilde{\Omega}_{N_{\rm CFPs}-1}}) & .& . & . \\  
    . & . & . & . & .\\ 
    . & . & . & . & . \\ 
     \psi^{n}({\Omega_1,\tilde{\Omega}_{1}}) & \psi^{n}({\Omega_2,\tilde{\Omega}_{1}})  & .& . & . \\  
    
	Y_{Q,l,-l}(\Omega_1) & Y_{Q,l,-l}(\Omega_2)  & . & . & . \\ 
        Y_{Q,l,-l+1}(\Omega_1) & Y_{Q,l,-l+1}(\Omega_2)  & . & . & . \\ 
	. & . & . & . & .\\ 
        Y_{Q,l,l}(\Omega_1) & Y_{Q,l,l}(\Omega_2)  & . & . & . \\ 
	. & . & . & . & . \\ 
        . & . & . & . & . \\ 
	\end{vmatrix}\nonumber \\
        & & \times \prod_{i<j}(u_iv_j-u_jv_i)^{2p},
    \label{equation:phiqh1/mdet}
\end{eqnarray}
where $l$ is the orbital angular momentum for the $n$th LL and $\tilde{\Omega}_{i}$ corresponds to the coordinates of the $i^{\rm th}$ CFP. The first $N_{\rm CFPs}$ rows are filled with coherent states in the $(n{+}1)$th $\Lambda$L [see Eq.~\eqref{eq: coherent state n LL}] to create CFPs, while the remaining rows are made up of single-particle eigenstates from the lowest $\Lambda$L to the $n$th $\Lambda$L.

As the CFP goes around in a closed loop, the initial wave function $\Psi_{I}$ under time evolution maps to a new wave function $\Psi_{F}{=} U \Psi_{I}$, where $U$ is a unitary evolution operator. For Abelian states, the initial and final wave functions differ only by a Berry phase, i.e., $U{=}e^ {i \xi}$. However, for non-Abelian states, the operator $U$ takes the form of a matrix that, along with incorporating Berry phase factors, transforms a set of original wave functions into a new set of wave functions. In this work, we focus on Jain states that are Abelian but we note that the technique we use was first utilized~\cite{Tserkovnyak03} to numerically demonstrate non-Abelian statistics of the quasiholes of the Moore-Read state~\cite{Moore91}. To compute the phase numerically, the CFP looping must be carried out in discrete steps by recursively calculating the evolution operator at every step. If the evolution operator at the $k^{\rm th}$ step is $U^k$, then the evolution operator at the next $(k{+}1)^{\rm th}$ step is given by
\begin{equation}
U^{k+1}=U^k\frac{[1-A^k/2]}{[1+A^k/2]},
\label{equation: numerical method}
\end{equation}
where $A$ is an anti-Hermitian operator defined as $A{=}\braket{\Psi|\dot{\Psi}}{=}(U^{-1} \dot{U})$~\cite{Wilczek83}. Provided the individual wave functions are normalized, $A$ can be redefined in a form that is most convenient for numerics as $A{=}[\braket{\Psi^k|\Psi^{k+1}}{-}\braket{\Psi^{k+1}|\Psi^{k}}]{/}2$, where $\Psi^{k}$ corresponds to the wave function at the $k^{\rm th}$ step, such that for $n_{\rm steps}$ steps $\Psi^{k=0}{=}\Psi_{I}$ and $\Psi^{k{=}n_{\rm steps}}{=}\Psi_{F}$. The numerical computations involving multi-dimensional integrals over the electron coordinates are carried out using the Metropolis Monte Carlo algorithm~\cite{Binder10}.

The braiding of one CFP around another fixed CFP is equivalent to rotating the entire spherical system about an axis passing through the fixed CFP. This rotational motion corresponds to calculating the system's angular momentum around the fixed CFP~\cite{Macaluso19}. Since the fixed CFP is placed at the north pole and the many-body wave function has azimuthal symmetry i.e., the wave function is an eigenstate of $L_{z}$, the phase accumulated at each step of the looping CFP is the same independent of its position on the loop (all the steps are of the same size). We have utilized this property to determine $U_1{=}e^{i\xi_1}$ at the first step and used it to find the total phase $\xi{=}n_{\rm steps} \xi_{1}$. This significantly reduces the computational cost of determining the evolution operator at multiple points of the trajectory of the looping CFP.

\section{Numerical Results}
\label{sec: numerical}
In this section, we present the numerical results of the Berry phase collected in the process of a CFP looping around a trajectory in the spherical geometry in the absence and presence of another CFP. We also present results on the statistics extracted using the numerical technique discussed in Sec.~\ref{subsec: extraction_of_statistics}.

\subsection{Looping of a single CFP}
We construct a many-body wave function of $N$ electrons with a single CFP at a filling fraction $n/(2pn{+}1)$ according to the CF theory as described in Eq.~\eqref{equation:phiqh1/mdet}. The directed (clockwise) motion of the CFP around a loop is implemented by moving its coordinates which are part of the coherent states that reside in the wave function. We loop a CFP at different latitudes that make different angles $\theta$ with the $z$-axis [see Fig.~\ref{fig: schematic_CFP_phase}(a)]. The entire trajectory of a single loop is broken into several discrete steps with the phase being updated at every step according to Eq.~\eqref{equation: numerical method} (the number of steps $n_{\rm steps}{\geq}100$ and is chosen sufficiently large such that the results are fully converged). The numerical value of the Berry phase at 1/3, 1/5, 2/5, 2/9, and 3/7 for a single CFP [for each fraction we show three system sizes] is shown in Fig.~\ref{fig: phase_qp_1}.  These results nicely fit the theoretical expectation (shown by thick lines in Fig.~\ref{fig: phase_qp_1}) given by Eq.~\eqref{eq: single qp phase}.

\begin{table}[t!]
    \begin{center}
        \begin{tabular}{ |c|c|c|c| } 
            \hline
            $\nu$ & $N$ & $\Theta$ \\
            \hline
            \multirow{3}{2em}{$1/3$} & $30$ & $0.6064(2)$  \\ 
            & $40$ & $0.6426(1)$  \\ 
            & $50$ & $0.6532(6)$  \\ 
            \hline
            \multirow{3}{2em}{$1/5$} & $30$ & $0.7143(2)$  \\ 
            & $40$ & $0.7762(2)$  \\ 
            & $50$ & $0.7970(5)$  \\ 
            \hline
            \multirow{3}{2em}{$2/5$} & $70$ & $0.3627(3)$  \\ 
            & $86$ & $0.3715(6)$  \\ 
            & $100$ & $0.3827(9)$  \\ 
            \hline
             \multirow{3}{2em}{$2/9$} & $70$ & $0.4059(4)$  \\ 
            & $86$ & $0.4166(4)$  \\ 
            & $100$ & $0.4369(5)$  \\ 
            \hline
            \multirow{3}{2em}{$3/7$} & $173$ & $0.3662(3)$  \\ 
            & $200$ & $0.3153(6)$ \\
            & $230$ & $0.2855(3)$ \\
           \hline
        \end{tabular}
    \end{center}
    \caption{Statistics parameter $\Theta$ extracted from the Berry phase accumulated when a CFP loops around a latitude in the presence of a CFP static at the north pole for various fillings $\nu$ in the Jain sequence for $N$ electrons in the spherical geometry. The parameter $\Theta$ is obtained by fitting the Berry phase to the form given in Eq.~\eqref{eq: mod_doub_qp_phase_final}. The number in the parenthesis is the error due to curve fitting.}
    \label{tab: frac_stat}
\end{table}

\begin{figure}[t!]

\subfloat{}{%
  \includegraphics[clip,width=0.78\columnwidth]{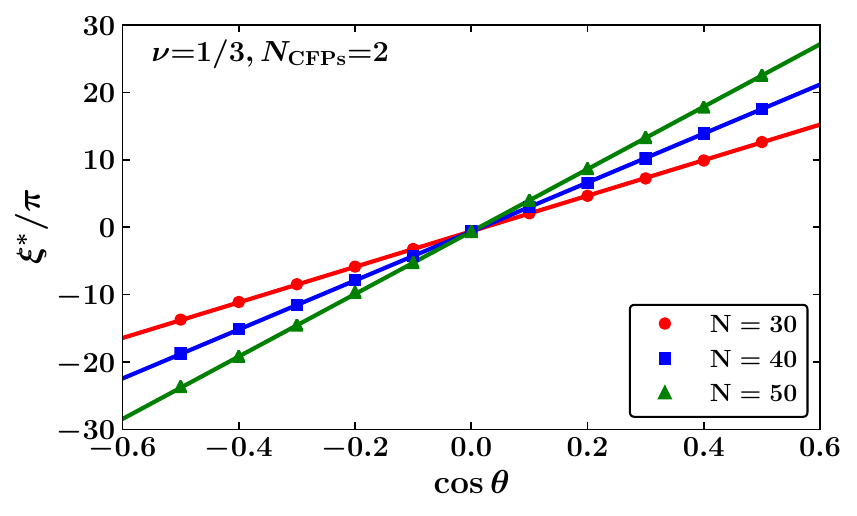}%
   
}

\subfloat{}{%
  \includegraphics[clip,width=0.78\columnwidth]{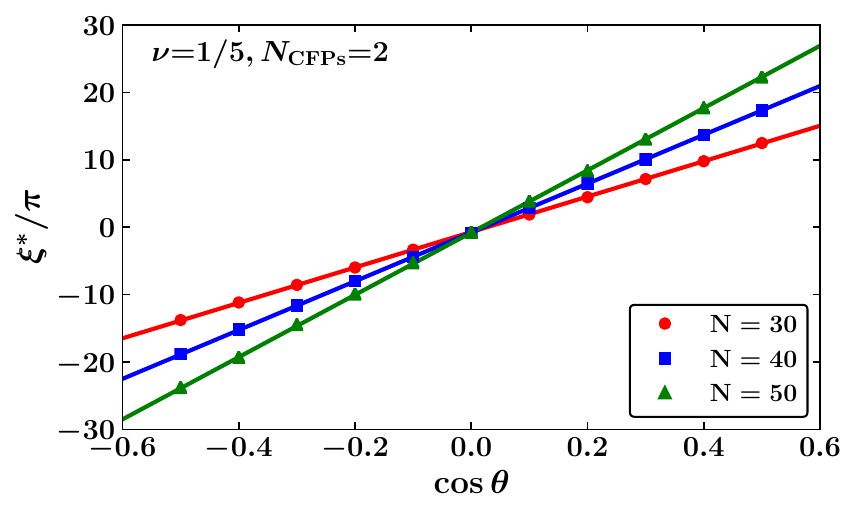}%
  
}

\subfloat{}{%
  \includegraphics[clip,width=0.78\columnwidth]{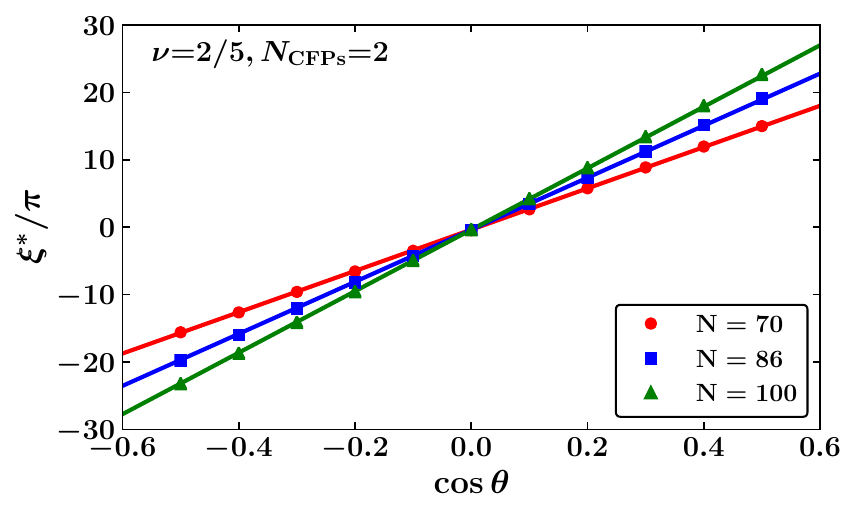}%
   
}

\subfloat{}{%
  \includegraphics[clip,width=0.78\columnwidth]{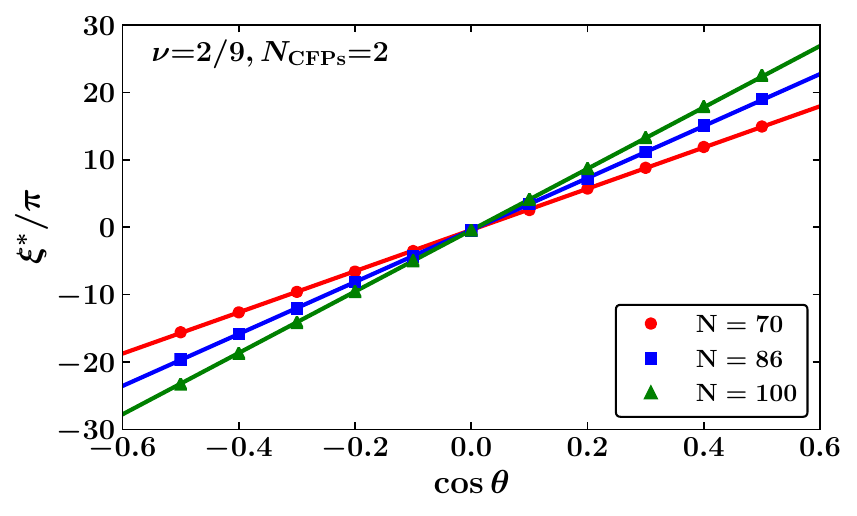}%
  
}

\subfloat{}{%
  \includegraphics[clip,width=0.78\columnwidth]{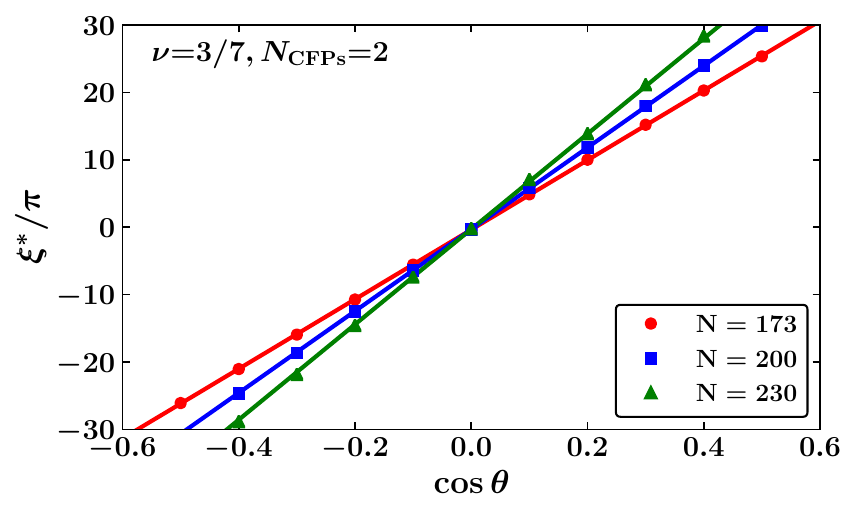}%
  
}

\caption{Phase accumulated due to a CFP looping along various latitudes $\theta$ in the presence of another CFP static at the north pole for multiple fillings. Solid dots and solid lines indicate numerical data and curve fitting using Eq.~\eqref{eq: mod_doub_qp_phase_final} respectively. }
\label{fig: phase_qp_2}
\end{figure}


\subsection{Looping of a CFP about another CFP}
Next, we examine a system consisting of two CFPs wherein one CFP is fixed at the north pole while the other CFP loops along a latitude $\theta$. As in the single CFP case, here we calculate the total phase accumulated by the looping operation and extract the braiding statistics by fitting it to Eq.~\eqref{eq: mod_doub_qp_phase_final} for filling fractions 1/3, 1/5, 2/5, 2/9, and 3/7 as shown in Fig.~\ref{fig: phase_qp_2}. The value of $\Theta$ extracted from the curve fitting, displayed in Table~\ref{tab: frac_stat}, is consistent with theoretical expectation of the statistics parameter [$\Theta{=}2p{/}(2pn{+}1)$] ~\cite{Wen95}. 

Due to the absence of an edge on the sphere, a bulk quantity such as the statistics, converges to the expected theoretical value even for relatively small systems. This enables us to extract an accurate value of $\Theta$ for $\nu{=}1{/}3{,}~1{/}5$ from systems as small as with $N{=}50$ electrons. However, as we try to access states with higher $\Lambda$Ls, i.e., higher $n$ in $n/(2pn{+}1)$, larger system sizes are necessary to avoid overlap between the CFPs, which are known to grow in size with increasing $n$~\cite{Johri14, Balram13b, Gattu23}. This is because the CFPs have a size that is the order of effective magnetic length $\ell^{*}{=}\sqrt{2pn{+}1}\ell$ and this grows with $n$. We note here that in many cases (results not presented here) we have checked that the phases evaluated using wave functions projected to the LLL result in the same statistics thereby supporting the hypothesis that topological properties are robust to operations like projection to the LLL~\cite{Balram16b, Anand22}.

\section{Discussion}
\label{sec: discussion}

In this work, we extracted the statistics of composite fermion particles in Jain FQH states using the spherical geometry. We showed that they respect Abelian statistics with the statistics parameter $\Theta{=}2p{/}(2pn{+}1)$ consistent with theoretical expectations~\cite{Wen95, Jain07}. We found that the shift ($\Delta$) in the phase arising from the shift in the position of a looping CFP due to the insertion of another CFP in the spherical geometry is equal to the shift ($\delta$) seen in the disk geometry. However, owing to the curvature, unlike in the planar geometry, the knowledge of the precise value of $\Delta$ is not necessary for the evaluation of statistics of a CFP in the spherical geometry. Additionally, the compact nature of the sphere facilitates obtaining converged results with much smaller systems compared to its disk counterparts.

Aside from the Jain states, many FQH states have been experimentally observed predominantly in the SLL~\cite{Willett87, Xia04, Kumar10} but some also in the LLL~\cite{Pan03, Samkharadze15b, Pan15, Kumar18} that fall outside the theory of free composite fermions. Many of these states can be understood as arising from residual interactions between the CFs~\cite{Sitko96, Read00, Scarola02, Mukherjee14, Balram15, Balram16c}. However, more recently, it has been shown that a generalization of the CF theory, known as the parton theory~\cite{Jain89b}, where FQH states are built as products of IQH states, can capture almost all the experimentally observed FQH states~\cite{Jain89b, Wu17, Balram18, Balram19, Faugno19, Balram20, Faugno20a, Balram20b, Faugno21, Balram21, Balram21a, Balram21b, Balram21c, Balram21d, Dora22, Sharma22, Bose23, Sharma23, Balram24}. The method we presented, which builds on IQH states, can be readily extended to extract statistics of the partons. In particular, excitations of certain partonic FQH states are expected to carry non-Abelian statistics~\cite{Wen91}. An interesting application of the techniques we presented would be to test for the presence of excitations with non-Abelian statistics using the microscopic parton wave functions. Recently, it has been shown that the statistics of a quasiparticle can also be discerned from a measurement of its density profile~\cite{Umucalilar18, Macaluso19, Macaluso20}. It would be useful to test the accuracy of the statistics of CFPs and CFHs ascertained from that method and compare and contrast it with the technique used in this work. We leave a detailed exploration of this and other ideas to future work.

\begin{acknowledgements} 
We acknowledge useful discussions with Abhimanyu Choudhury, Moty Heiblum, Jainendra K. Jain, and Steven H. Simon. Computational portions of this research work were conducted using the Nandadevi supercomputer, which is maintained and supported by the Institute of Mathematical Science's High-Performance Computing Center. ACB thanks the Science and Engineering Research Board (SERB) of the Department of Science and Technology (DST) for funding support via the Mathematical Research Impact Centric Support (MATRICS) Grant No. MTR/2023/000002.
\end{acknowledgements}
		
\bibliography{biblio_fqhe}

\begin{thebibliography}{76}%
\makeatletter
\providecommand \@ifxundefined [1]{%
 \@ifx{#1\undefined}
}%
\providecommand \@ifnum [1]{%
 \ifnum #1\expandafter \@firstoftwo
 \else \expandafter \@secondoftwo
 \fi
}%
\providecommand \@ifx [1]{%
 \ifx #1\expandafter \@firstoftwo
 \else \expandafter \@secondoftwo
 \fi
}%
\providecommand \natexlab [1]{#1}%
\providecommand \enquote  [1]{``#1''}%
\providecommand \bibnamefont  [1]{#1}%
\providecommand \bibfnamefont [1]{#1}%
\providecommand \citenamefont [1]{#1}%
\providecommand \href@noop [0]{\@secondoftwo}%
\providecommand \href [0]{\begingroup \@sanitize@url \@href}%
\providecommand \@href[1]{\@@startlink{#1}\@@href}%
\providecommand \@@href[1]{\endgroup#1\@@endlink}%
\providecommand \@sanitize@url [0]{\catcode `\\12\catcode `\$12\catcode
  `\&12\catcode `\#12\catcode `\^12\catcode `\_12\catcode `\%12\relax}%
\providecommand \@@startlink[1]{}%
\providecommand \@@endlink[0]{}%
\providecommand \url  [0]{\begingroup\@sanitize@url \@url }%
\providecommand \@url [1]{\endgroup\@href {#1}{\urlprefix }}%
\providecommand \urlprefix  [0]{URL }%
\providecommand \Eprint [0]{\href }%
\providecommand \doibase [0]{https://doi.org/}%
\providecommand \selectlanguage [0]{\@gobble}%
\providecommand \bibinfo  [0]{\@secondoftwo}%
\providecommand \bibfield  [0]{\@secondoftwo}%
\providecommand \translation [1]{[#1]}%
\providecommand \BibitemOpen [0]{}%
\providecommand \bibitemStop [0]{}%
\providecommand \bibitemNoStop [0]{.\EOS\space}%
\providecommand \EOS [0]{\spacefactor3000\relax}%
\providecommand \BibitemShut  [1]{\csname bibitem#1\endcsname}%
\let\auto@bib@innerbib\@empty
\bibitem [{\citenamefont {Leinaas}\ and\ \citenamefont
  {Myrheim}(1977)}]{Leinaas77}%
  \BibitemOpen
  \bibfield  {author} {\bibinfo {author} {\bibfnamefont {J.}~\bibnamefont
  {Leinaas}}\ and\ \bibinfo {author} {\bibfnamefont {J.}~\bibnamefont
  {Myrheim}},\ }\bibfield  {title} {\bibinfo {title} {On the theory of
  identical particles},\ }\href {https://doi.org/10.1007/BF02727953} {\bibfield
   {journal} {\bibinfo  {journal} {Il Nuovo Cimento B Series 11}\ }\textbf
  {\bibinfo {volume} {37}},\ \bibinfo {pages} {1} (\bibinfo {year}
  {1977})}\BibitemShut {NoStop}%
\bibitem [{\citenamefont {Tsui}\ \emph {et~al.}(1982)\citenamefont {Tsui},
  \citenamefont {Stormer},\ and\ \citenamefont {Gossard}}]{Tsui82}%
  \BibitemOpen
  \bibfield  {author} {\bibinfo {author} {\bibfnamefont {D.~C.}\ \bibnamefont
  {Tsui}}, \bibinfo {author} {\bibfnamefont {H.~L.}\ \bibnamefont {Stormer}},\
  and\ \bibinfo {author} {\bibfnamefont {A.~C.}\ \bibnamefont {Gossard}},\
  }\bibfield  {title} {\bibinfo {title} {Two-dimensional magnetotransport in
  the extreme quantum limit},\ }\href
  {https://doi.org/10.1103/PhysRevLett.48.1559} {\bibfield  {journal} {\bibinfo
   {journal} {Phys. Rev. Lett.}\ }\textbf {\bibinfo {volume} {48}},\ \bibinfo
  {pages} {1559} (\bibinfo {year} {1982})}\BibitemShut {NoStop}%
\bibitem [{\citenamefont {Laughlin}(1983)}]{Laughlin83}%
  \BibitemOpen
  \bibfield  {author} {\bibinfo {author} {\bibfnamefont {R.~B.}\ \bibnamefont
  {Laughlin}},\ }\bibfield  {title} {\bibinfo {title} {Anomalous quantum {Hall}
  effect: An incompressible quantum fluid with fractionally charged
  excitations},\ }\href {https://doi.org/10.1103/PhysRevLett.50.1395}
  {\bibfield  {journal} {\bibinfo  {journal} {Phys. Rev. Lett.}\ }\textbf
  {\bibinfo {volume} {50}},\ \bibinfo {pages} {1395} (\bibinfo {year}
  {1983})}\BibitemShut {NoStop}%
\bibitem [{\citenamefont {Wilczek}(1982)}]{Wilczek82}%
  \BibitemOpen
  \bibfield  {author} {\bibinfo {author} {\bibfnamefont {F.}~\bibnamefont
  {Wilczek}},\ }\bibfield  {title} {\bibinfo {title} {Quantum mechanics of
  fractional-spin particles},\ }\href
  {https://doi.org/10.1103/PhysRevLett.49.957} {\bibfield  {journal} {\bibinfo
  {journal} {Phys. Rev. Lett.}\ }\textbf {\bibinfo {volume} {49}},\ \bibinfo
  {pages} {957} (\bibinfo {year} {1982})}\BibitemShut {NoStop}%
\bibitem [{\citenamefont {Arovas}\ \emph {et~al.}(1984)\citenamefont {Arovas},
  \citenamefont {Schrieffer},\ and\ \citenamefont {Wilczek}}]{Arovas84}%
  \BibitemOpen
  \bibfield  {author} {\bibinfo {author} {\bibfnamefont {D.}~\bibnamefont
  {Arovas}}, \bibinfo {author} {\bibfnamefont {J.~R.}\ \bibnamefont
  {Schrieffer}},\ and\ \bibinfo {author} {\bibfnamefont {F.}~\bibnamefont
  {Wilczek}},\ }\bibfield  {title} {\bibinfo {title} {Fractional statistics and
  the quantum {Hall} effect},\ }\href
  {https://doi.org/10.1103/PhysRevLett.53.722} {\bibfield  {journal} {\bibinfo
  {journal} {Phys. Rev. Lett.}\ }\textbf {\bibinfo {volume} {53}},\ \bibinfo
  {pages} {722} (\bibinfo {year} {1984})}\BibitemShut {NoStop}%
\bibitem [{\citenamefont {Halperin}(1984)}]{Halperin84}%
  \BibitemOpen
  \bibfield  {author} {\bibinfo {author} {\bibfnamefont {B.~I.}\ \bibnamefont
  {Halperin}},\ }\bibfield  {title} {\bibinfo {title} {Statistics of
  quasiparticles and the hierarchy of fractional quantized {Hall} states},\
  }\href {https://doi.org/10.1103/PhysRevLett.52.1583} {\bibfield  {journal}
  {\bibinfo  {journal} {Phys. Rev. Lett.}\ }\textbf {\bibinfo {volume} {52}},\
  \bibinfo {pages} {1583} (\bibinfo {year} {1984})}\BibitemShut {NoStop}%
\bibitem [{\citenamefont {de~Picciotto}\ \emph {et~al.}(1997)\citenamefont
  {de~Picciotto}, \citenamefont {Reznikov}, \citenamefont {Heiblum},
  \citenamefont {Umansky}, \citenamefont {Bunin},\ and\ \citenamefont
  {Mahalu}}]{de-Picciotto97}%
  \BibitemOpen
  \bibfield  {author} {\bibinfo {author} {\bibfnamefont {R.}~\bibnamefont
  {de~Picciotto}}, \bibinfo {author} {\bibfnamefont {M.}~\bibnamefont
  {Reznikov}}, \bibinfo {author} {\bibfnamefont {M.}~\bibnamefont {Heiblum}},
  \bibinfo {author} {\bibfnamefont {V.}~\bibnamefont {Umansky}}, \bibinfo
  {author} {\bibfnamefont {G.}~\bibnamefont {Bunin}},\ and\ \bibinfo {author}
  {\bibfnamefont {D.}~\bibnamefont {Mahalu}},\ }\bibfield  {title} {\bibinfo
  {title} {Direct observation of a fractional charge},\ }\href@noop {}
  {\bibfield  {journal} {\bibinfo  {journal} {Nature}\ }\textbf {\bibinfo
  {volume} {389}},\ \bibinfo {pages} {162} (\bibinfo {year}
  {1997})}\BibitemShut {NoStop}%
\bibitem [{\citenamefont {Nakamura}\ \emph {et~al.}(2020)\citenamefont
  {Nakamura}, \citenamefont {Liang}, \citenamefont {Gardner},\ and\
  \citenamefont {Manfra}}]{Nakamura20}%
  \BibitemOpen
  \bibfield  {author} {\bibinfo {author} {\bibfnamefont {J.}~\bibnamefont
  {Nakamura}}, \bibinfo {author} {\bibfnamefont {S.}~\bibnamefont {Liang}},
  \bibinfo {author} {\bibfnamefont {G.~C.}\ \bibnamefont {Gardner}},\ and\
  \bibinfo {author} {\bibfnamefont {M.~J.}\ \bibnamefont {Manfra}},\ }\bibfield
   {title} {\bibinfo {title} {Direct observation of anyonic braiding
  statistics},\ }\href@noop {} {\bibfield  {journal} {\bibinfo  {journal}
  {Nature Physics}\ }\textbf {\bibinfo {volume} {16}},\ \bibinfo {pages} {931}
  (\bibinfo {year} {2020})}\BibitemShut {NoStop}%
\bibitem [{\citenamefont {Bartolomei}\ \emph {et~al.}(2020)\citenamefont
  {Bartolomei}, \citenamefont {Kumar}, \citenamefont {Bisognin}, \citenamefont
  {Marguerite}, \citenamefont {Berroir}, \citenamefont {Bocquillon},
  \citenamefont {Pla{\c c}ais}, \citenamefont {Cavanna}, \citenamefont {Dong},
  \citenamefont {Gennser}, \citenamefont {Jin},\ and\ \citenamefont
  {F{\`e}ve}}]{Bartolomei20}%
  \BibitemOpen
  \bibfield  {author} {\bibinfo {author} {\bibfnamefont {H.}~\bibnamefont
  {Bartolomei}}, \bibinfo {author} {\bibfnamefont {M.}~\bibnamefont {Kumar}},
  \bibinfo {author} {\bibfnamefont {R.}~\bibnamefont {Bisognin}}, \bibinfo
  {author} {\bibfnamefont {A.}~\bibnamefont {Marguerite}}, \bibinfo {author}
  {\bibfnamefont {J.-M.}\ \bibnamefont {Berroir}}, \bibinfo {author}
  {\bibfnamefont {E.}~\bibnamefont {Bocquillon}}, \bibinfo {author}
  {\bibfnamefont {B.}~\bibnamefont {Pla{\c c}ais}}, \bibinfo {author}
  {\bibfnamefont {A.}~\bibnamefont {Cavanna}}, \bibinfo {author} {\bibfnamefont
  {Q.}~\bibnamefont {Dong}}, \bibinfo {author} {\bibfnamefont {U.}~\bibnamefont
  {Gennser}}, \bibinfo {author} {\bibfnamefont {Y.}~\bibnamefont {Jin}},\ and\
  \bibinfo {author} {\bibfnamefont {G.}~\bibnamefont {F{\`e}ve}},\ }\bibfield
  {title} {\bibinfo {title} {Fractional statistics in anyon collisions},\
  }\href {https://doi.org/10.1126/science.aaz5601} {\bibfield  {journal}
  {\bibinfo  {journal} {Science}\ }\textbf {\bibinfo {volume} {368}},\ \bibinfo
  {pages} {173} (\bibinfo {year} {2020})},\ \Eprint
  {https://arxiv.org/abs/https://science.sciencemag.org/content/368/6487/173.full.pdf}
  {https://science.sciencemag.org/content/368/6487/173.full.pdf} \BibitemShut
  {NoStop}%
\bibitem [{\citenamefont {Nakamura}\ \emph {et~al.}(2023)\citenamefont
  {Nakamura}, \citenamefont {Liang}, \citenamefont {Gardner},\ and\
  \citenamefont {Manfra}}]{Nakamura23}%
  \BibitemOpen
  \bibfield  {author} {\bibinfo {author} {\bibfnamefont {J.}~\bibnamefont
  {Nakamura}}, \bibinfo {author} {\bibfnamefont {S.}~\bibnamefont {Liang}},
  \bibinfo {author} {\bibfnamefont {G.~C.}\ \bibnamefont {Gardner}},\ and\
  \bibinfo {author} {\bibfnamefont {M.~J.}\ \bibnamefont {Manfra}},\ }\bibfield
   {title} {\bibinfo {title} {{Fabry}-{P\'erot} interferometry at the
  $\ensuremath{\nu}=2/5$ fractional quantum {Hall} state},\ }\href
  {https://doi.org/10.1103/PhysRevX.13.041012} {\bibfield  {journal} {\bibinfo
  {journal} {Phys. Rev. X}\ }\textbf {\bibinfo {volume} {13}},\ \bibinfo
  {pages} {041012} (\bibinfo {year} {2023})}\BibitemShut {NoStop}%
\bibitem [{\citenamefont {Kim}\ \emph {et~al.}(2024)\citenamefont {Kim},
  \citenamefont {Dev}, \citenamefont {Kumar}, \citenamefont {Ilin},
  \citenamefont {Haug}, \citenamefont {Bhardwaj}, \citenamefont {Hong},
  \citenamefont {Watanabe}, \citenamefont {Taniguchi}, \citenamefont {Stern},\
  and\ \citenamefont {Ronen}}]{kim2024aharonovbohm}%
  \BibitemOpen
  \bibfield  {author} {\bibinfo {author} {\bibfnamefont {J.}~\bibnamefont
  {Kim}}, \bibinfo {author} {\bibfnamefont {H.}~\bibnamefont {Dev}}, \bibinfo
  {author} {\bibfnamefont {R.}~\bibnamefont {Kumar}}, \bibinfo {author}
  {\bibfnamefont {A.}~\bibnamefont {Ilin}}, \bibinfo {author} {\bibfnamefont
  {A.}~\bibnamefont {Haug}}, \bibinfo {author} {\bibfnamefont {V.}~\bibnamefont
  {Bhardwaj}}, \bibinfo {author} {\bibfnamefont {C.}~\bibnamefont {Hong}},
  \bibinfo {author} {\bibfnamefont {K.}~\bibnamefont {Watanabe}}, \bibinfo
  {author} {\bibfnamefont {T.}~\bibnamefont {Taniguchi}}, \bibinfo {author}
  {\bibfnamefont {A.}~\bibnamefont {Stern}},\ and\ \bibinfo {author}
  {\bibfnamefont {Y.}~\bibnamefont {Ronen}},\ }\href@noop {} {\bibinfo {title}
  {{Aharonov}-{Bohm} interference and the evolution of phase jumps in
  fractional quantum {Hall} {Fabry}-{Perot} interferometers based on bi-layer
  graphene}} (\bibinfo {year} {2024}),\ \Eprint
  {https://arxiv.org/abs/2402.12432} {arXiv:2402.12432 [cond-mat.mes-hall]}
  \BibitemShut {NoStop}%
\bibitem [{\citenamefont {Samuelson}\ \emph {et~al.}(2024)\citenamefont
  {Samuelson}, \citenamefont {Cohen}, \citenamefont {Wang}, \citenamefont
  {Blanch}, \citenamefont {Taniguchi}, \citenamefont {Watanabe}, \citenamefont
  {Zaletel},\ and\ \citenamefont {Young}}]{Samuelson24}%
  \BibitemOpen
  \bibfield  {author} {\bibinfo {author} {\bibfnamefont {N.~L.}\ \bibnamefont
  {Samuelson}}, \bibinfo {author} {\bibfnamefont {L.~A.}\ \bibnamefont
  {Cohen}}, \bibinfo {author} {\bibfnamefont {W.}~\bibnamefont {Wang}},
  \bibinfo {author} {\bibfnamefont {S.}~\bibnamefont {Blanch}}, \bibinfo
  {author} {\bibfnamefont {T.}~\bibnamefont {Taniguchi}}, \bibinfo {author}
  {\bibfnamefont {K.}~\bibnamefont {Watanabe}}, \bibinfo {author}
  {\bibfnamefont {M.~P.}\ \bibnamefont {Zaletel}},\ and\ \bibinfo {author}
  {\bibfnamefont {A.~F.}\ \bibnamefont {Young}},\ }\href@noop {} {\bibinfo
  {title} {Anyonic statistics and slow quasiparticle dynamics in a graphene
  fractional quantum {Hall} interferometer}} (\bibinfo {year} {2024}),\ \Eprint
  {https://arxiv.org/abs/2403.19628} {arXiv:2403.19628 [cond-mat.mes-hall]}
  \BibitemShut {NoStop}%
\bibitem [{\citenamefont {Werkmeister}\ \emph {et~al.}(2024)\citenamefont
  {Werkmeister}, \citenamefont {Ehrets}, \citenamefont {Wesson}, \citenamefont
  {Najafabadi}, \citenamefont {Watanabe}, \citenamefont {Taniguchi},
  \citenamefont {Halperin}, \citenamefont {Yacoby},\ and\ \citenamefont
  {Kim}}]{Werkmeister24}%
  \BibitemOpen
  \bibfield  {author} {\bibinfo {author} {\bibfnamefont {T.}~\bibnamefont
  {Werkmeister}}, \bibinfo {author} {\bibfnamefont {J.~R.}\ \bibnamefont
  {Ehrets}}, \bibinfo {author} {\bibfnamefont {M.~E.}\ \bibnamefont {Wesson}},
  \bibinfo {author} {\bibfnamefont {D.~H.}\ \bibnamefont {Najafabadi}},
  \bibinfo {author} {\bibfnamefont {K.}~\bibnamefont {Watanabe}}, \bibinfo
  {author} {\bibfnamefont {T.}~\bibnamefont {Taniguchi}}, \bibinfo {author}
  {\bibfnamefont {B.~I.}\ \bibnamefont {Halperin}}, \bibinfo {author}
  {\bibfnamefont {A.}~\bibnamefont {Yacoby}},\ and\ \bibinfo {author}
  {\bibfnamefont {P.}~\bibnamefont {Kim}},\ }\href@noop {} {\bibinfo {title}
  {Anyon braiding and telegraph noise in a graphene interferometer}} (\bibinfo
  {year} {2024}),\ \Eprint {https://arxiv.org/abs/2403.18983} {arXiv:2403.18983
  [cond-mat.mes-hall]} \BibitemShut {NoStop}%
\bibitem [{\citenamefont {Read}\ and\ \citenamefont
  {Das~Sarma}(2023)}]{Read23}%
  \BibitemOpen
  \bibfield  {author} {\bibinfo {author} {\bibfnamefont {N.}~\bibnamefont
  {Read}}\ and\ \bibinfo {author} {\bibfnamefont {S.}~\bibnamefont
  {Das~Sarma}},\ }\bibfield  {title} {\bibinfo {title} {Clarification of
  braiding statistics in {Fabry}-{Perot} interferometry},\ }\bibfield
  {journal} {\bibinfo  {journal} {Nature Physics}\ }\href
  {https://doi.org/10.1038/s41567-023-02309-8} {10.1038/s41567-023-02309-8}
  (\bibinfo {year} {2023})\BibitemShut {NoStop}%
\bibitem [{\citenamefont {Kivelson}\ and\ \citenamefont
  {Murthy}(2024)}]{Kivelson24}%
  \BibitemOpen
  \bibfield  {author} {\bibinfo {author} {\bibfnamefont {S.~A.}\ \bibnamefont
  {Kivelson}}\ and\ \bibinfo {author} {\bibfnamefont {C.}~\bibnamefont
  {Murthy}},\ }\href@noop {} {\bibinfo {title} {A modified interferometer to
  measure anyonic braiding statistics}} (\bibinfo {year} {2024}),\ \Eprint
  {https://arxiv.org/abs/2403.12139} {arXiv:2403.12139 [cond-mat.mes-hall]}
  \BibitemShut {NoStop}%
\bibitem [{\citenamefont {Freedman}\ \emph {et~al.}(2003)\citenamefont
  {Freedman}, \citenamefont {Kitaev}, \citenamefont {Larsen},\ and\
  \citenamefont {Wang}}]{Freedman03}%
  \BibitemOpen
  \bibfield  {author} {\bibinfo {author} {\bibfnamefont {M.}~\bibnamefont
  {Freedman}}, \bibinfo {author} {\bibfnamefont {A.}~\bibnamefont {Kitaev}},
  \bibinfo {author} {\bibfnamefont {M.}~\bibnamefont {Larsen}},\ and\ \bibinfo
  {author} {\bibfnamefont {Z.}~\bibnamefont {Wang}},\ }\bibfield  {title}
  {\bibinfo {title} {Topological quantum computation},\ }\href
  {https://www.ams.org/journals/bull/2003-40-01/S0273-0979-02-00964-3}
  {\bibfield  {journal} {\bibinfo  {journal} {Bulletin of the American
  Mathematical Society}\ }\textbf {\bibinfo {volume} {40}},\ \bibinfo {pages}
  {31} (\bibinfo {year} {2003})}\BibitemShut {NoStop}%
\bibitem [{\citenamefont {Eisenstein}\ and\ \citenamefont
  {Stormer}(1990)}]{Eisenstein90a}%
  \BibitemOpen
  \bibfield  {author} {\bibinfo {author} {\bibfnamefont {J.~P.}\ \bibnamefont
  {Eisenstein}}\ and\ \bibinfo {author} {\bibfnamefont {H.~L.}\ \bibnamefont
  {Stormer}},\ }\bibfield  {title} {\bibinfo {title} {The fractional quantum
  {Hall} effect},\ }\href {https://doi.org/10.1126/science.248.4962.1510}
  {\bibfield  {journal} {\bibinfo  {journal} {Science}\ }\textbf {\bibinfo
  {volume} {248}},\ \bibinfo {pages} {1510} (\bibinfo {year} {1990})},\ \Eprint
  {https://arxiv.org/abs/https://www.science.org/doi/pdf/10.1126/science.248.4962.1510}
  {https://www.science.org/doi/pdf/10.1126/science.248.4962.1510} \BibitemShut
  {NoStop}%
\bibitem [{\citenamefont {Jain}(1989{\natexlab{a}})}]{Jain89}%
  \BibitemOpen
  \bibfield  {author} {\bibinfo {author} {\bibfnamefont {J.~K.}\ \bibnamefont
  {Jain}},\ }\bibfield  {title} {\bibinfo {title} {Composite-fermion approach
  for the fractional quantum {Hall} effect},\ }\href
  {https://doi.org/10.1103/PhysRevLett.63.199} {\bibfield  {journal} {\bibinfo
  {journal} {Phys. Rev. Lett.}\ }\textbf {\bibinfo {volume} {63}},\ \bibinfo
  {pages} {199} (\bibinfo {year} {1989}{\natexlab{a}})}\BibitemShut {NoStop}%
\bibitem [{\citenamefont {Jain}(2007)}]{Jain07}%
  \BibitemOpen
  \bibfield  {author} {\bibinfo {author} {\bibfnamefont {J.~K.}\ \bibnamefont
  {Jain}},\ }\href@noop {} {\emph {\bibinfo {title} {Composite Fermions}}}\
  (\bibinfo  {publisher} {Cambridge University Press, New York, US},\ \bibinfo
  {year} {2007})\BibitemShut {NoStop}%
\bibitem [{\citenamefont {Kj{\o}nsberg}\ and\ \citenamefont
  {Myrheim}(1999)}]{Kjonsberg99}%
  \BibitemOpen
  \bibfield  {author} {\bibinfo {author} {\bibfnamefont {H.}~\bibnamefont
  {Kj{\o}nsberg}}\ and\ \bibinfo {author} {\bibfnamefont {J.}~\bibnamefont
  {Myrheim}},\ }\bibfield  {title} {\bibinfo {title} {Numerical study of charge
  and statistics of {Laughlin} quasiparticles},\ }\href
  {https://doi.org/10.1142/S0217751X99000270} {\bibfield  {journal} {\bibinfo
  {journal} {International Journal of Modern Physics A}\ }\textbf {\bibinfo
  {volume} {14}},\ \bibinfo {pages} {537} (\bibinfo {year} {1999})}\BibitemShut
  {NoStop}%
\bibitem [{\citenamefont {Nardin}\ \emph {et~al.}(2023)\citenamefont {Nardin},
  \citenamefont {Ardonne},\ and\ \citenamefont {Mazza}}]{Nardin23}%
  \BibitemOpen
  \bibfield  {author} {\bibinfo {author} {\bibfnamefont {A.}~\bibnamefont
  {Nardin}}, \bibinfo {author} {\bibfnamefont {E.}~\bibnamefont {Ardonne}},\
  and\ \bibinfo {author} {\bibfnamefont {L.}~\bibnamefont {Mazza}},\ }\bibfield
   {title} {\bibinfo {title} {Spin-statistics relation for quantum {Hall}
  states},\ }\href {https://doi.org/10.1103/PhysRevB.108.L041105} {\bibfield
  {journal} {\bibinfo  {journal} {Phys. Rev. B}\ }\textbf {\bibinfo {volume}
  {108}},\ \bibinfo {pages} {L041105} (\bibinfo {year} {2023})}\BibitemShut
  {NoStop}%
\bibitem [{\citenamefont {Kj{\o}nsberg}\ and\ \citenamefont
  {Leinaas}(1999)}]{Kjonsberg99b}%
  \BibitemOpen
  \bibfield  {author} {\bibinfo {author} {\bibfnamefont {H.}~\bibnamefont
  {Kj{\o}nsberg}}\ and\ \bibinfo {author} {\bibfnamefont {J.~M.}\ \bibnamefont
  {Leinaas}},\ }\bibfield  {title} {\bibinfo {title} {Charge and statistics of
  quantum {Hall} quasi-particles. a numerical study of mean values and
  fluctuations},\ }\href {https://doi.org/10.1016/S0550-3213(99)00353-3}
  {\bibfield  {journal} {\bibinfo  {journal} {Nucl. Phys. B}\ }\textbf
  {\bibinfo {volume} {559}},\ \bibinfo {pages} {705} (\bibinfo {year}
  {1999})}\BibitemShut {NoStop}%
\bibitem [{\citenamefont {Jeon}\ \emph {et~al.}(2004)\citenamefont {Jeon},
  \citenamefont {Graham},\ and\ \citenamefont {Jain}}]{Jeon04}%
  \BibitemOpen
  \bibfield  {author} {\bibinfo {author} {\bibfnamefont {G.~S.}\ \bibnamefont
  {Jeon}}, \bibinfo {author} {\bibfnamefont {K.~L.}\ \bibnamefont {Graham}},\
  and\ \bibinfo {author} {\bibfnamefont {J.~K.}\ \bibnamefont {Jain}},\
  }\bibfield  {title} {\bibinfo {title} {Berry phases for composite fermions:
  Effective magnetic field and fractional statistics},\ }\href
  {https://doi.org/10.1103/PhysRevB.70.125316} {\bibfield  {journal} {\bibinfo
  {journal} {Phys. Rev. B}\ }\textbf {\bibinfo {volume} {70}},\ \bibinfo
  {pages} {125316} (\bibinfo {year} {2004})}\BibitemShut {NoStop}%
\bibitem [{\citenamefont {Tserkovnyak}\ and\ \citenamefont
  {Simon}(2003)}]{Tserkovnyak03}%
  \BibitemOpen
  \bibfield  {author} {\bibinfo {author} {\bibfnamefont {Y.}~\bibnamefont
  {Tserkovnyak}}\ and\ \bibinfo {author} {\bibfnamefont {S.~H.}\ \bibnamefont
  {Simon}},\ }\bibfield  {title} {\bibinfo {title} {{Monte} {Carlo} evaluation
  of non-abelian statistics},\ }\href
  {https://doi.org/10.1103/PhysRevLett.90.016802} {\bibfield  {journal}
  {\bibinfo  {journal} {Phys. Rev. Lett.}\ }\textbf {\bibinfo {volume} {90}},\
  \bibinfo {pages} {016802} (\bibinfo {year} {2003})}\BibitemShut {NoStop}%
\bibitem [{\citenamefont {Balram}\ \emph
  {et~al.}(2013{\natexlab{a}})\citenamefont {Balram}, \citenamefont {Wu},
  \citenamefont {Sreejith}, \citenamefont {W\'ojs},\ and\ \citenamefont
  {Jain}}]{Balram13b}%
  \BibitemOpen
  \bibfield  {author} {\bibinfo {author} {\bibfnamefont {A.~C.}\ \bibnamefont
  {Balram}}, \bibinfo {author} {\bibfnamefont {Y.-H.}\ \bibnamefont {Wu}},
  \bibinfo {author} {\bibfnamefont {G.~J.}\ \bibnamefont {Sreejith}}, \bibinfo
  {author} {\bibfnamefont {A.}~\bibnamefont {W\'ojs}},\ and\ \bibinfo {author}
  {\bibfnamefont {J.~K.}\ \bibnamefont {Jain}},\ }\bibfield  {title} {\bibinfo
  {title} {Role of exciton screening in the $\nu=7/3$ fractional quantum {Hall}
  effect},\ }\href {https://doi.org/10.1103/PhysRevLett.110.186801} {\bibfield
  {journal} {\bibinfo  {journal} {Phys. Rev. Lett.}\ }\textbf {\bibinfo
  {volume} {110}},\ \bibinfo {pages} {186801} (\bibinfo {year}
  {2013}{\natexlab{a}})}\BibitemShut {NoStop}%
\bibitem [{\citenamefont {Balram}\ and\ \citenamefont
  {Jain}(2016)}]{Balram16b}%
  \BibitemOpen
  \bibfield  {author} {\bibinfo {author} {\bibfnamefont {A.~C.}\ \bibnamefont
  {Balram}}\ and\ \bibinfo {author} {\bibfnamefont {J.~K.}\ \bibnamefont
  {Jain}},\ }\bibfield  {title} {\bibinfo {title} {Nature of composite fermions
  and the role of particle-hole symmetry: A microscopic account},\ }\href
  {https://doi.org/10.1103/PhysRevB.93.235152} {\bibfield  {journal} {\bibinfo
  {journal} {Phys. Rev. B}\ }\textbf {\bibinfo {volume} {93}},\ \bibinfo
  {pages} {235152} (\bibinfo {year} {2016})}\BibitemShut {NoStop}%
\bibitem [{\citenamefont {Anand}\ \emph {et~al.}(2022)\citenamefont {Anand},
  \citenamefont {Patil}, \citenamefont {Balram},\ and\ \citenamefont
  {Sreejith}}]{Anand22}%
  \BibitemOpen
  \bibfield  {author} {\bibinfo {author} {\bibfnamefont {A.}~\bibnamefont
  {Anand}}, \bibinfo {author} {\bibfnamefont {R.~A.}\ \bibnamefont {Patil}},
  \bibinfo {author} {\bibfnamefont {A.~C.}\ \bibnamefont {Balram}},\ and\
  \bibinfo {author} {\bibfnamefont {G.~J.}\ \bibnamefont {Sreejith}},\
  }\bibfield  {title} {\bibinfo {title} {Real-space entanglement spectra of
  parton states in fractional quantum {Hall} systems},\ }\href
  {https://doi.org/10.1103/PhysRevB.106.085136} {\bibfield  {journal} {\bibinfo
   {journal} {Phys. Rev. B}\ }\textbf {\bibinfo {volume} {106}},\ \bibinfo
  {pages} {085136} (\bibinfo {year} {2022})}\BibitemShut {NoStop}%
\bibitem [{\citenamefont {Haldane}(1983)}]{Haldane83}%
  \BibitemOpen
  \bibfield  {author} {\bibinfo {author} {\bibfnamefont {F.~D.~M.}\
  \bibnamefont {Haldane}},\ }\bibfield  {title} {\bibinfo {title} {Fractional
  quantization of the {Hall} effect: A hierarchy of incompressible quantum
  fluid states},\ }\href {https://doi.org/10.1103/PhysRevLett.51.605}
  {\bibfield  {journal} {\bibinfo  {journal} {Phys. Rev. Lett.}\ }\textbf
  {\bibinfo {volume} {51}},\ \bibinfo {pages} {605} (\bibinfo {year}
  {1983})}\BibitemShut {NoStop}%
\bibitem [{\citenamefont {Wen}\ and\ \citenamefont {Zee}(1992)}]{Wen92}%
  \BibitemOpen
  \bibfield  {author} {\bibinfo {author} {\bibfnamefont {X.~G.}\ \bibnamefont
  {Wen}}\ and\ \bibinfo {author} {\bibfnamefont {A.}~\bibnamefont {Zee}},\
  }\bibfield  {title} {\bibinfo {title} {Shift and spin vector: New topological
  quantum numbers for the {Hall} fluids},\ }\href
  {https://doi.org/10.1103/PhysRevLett.69.953} {\bibfield  {journal} {\bibinfo
  {journal} {Phys. Rev. Lett.}\ }\textbf {\bibinfo {volume} {69}},\ \bibinfo
  {pages} {953} (\bibinfo {year} {1992})}\BibitemShut {NoStop}%
\bibitem [{\citenamefont {Wu}\ and\ \citenamefont {Yang}(1976)}]{Wu76}%
  \BibitemOpen
  \bibfield  {author} {\bibinfo {author} {\bibfnamefont {T.~T.}\ \bibnamefont
  {Wu}}\ and\ \bibinfo {author} {\bibfnamefont {C.~N.}\ \bibnamefont {Yang}},\
  }\bibfield  {title} {\bibinfo {title} {Dirac monopole without strings:
  Monopole harmonics},\ }\href {https://doi.org/10.1016/0550-3213(76)90143-7}
  {\bibfield  {journal} {\bibinfo  {journal} {Nucl. Phys. B}\ }\textbf
  {\bibinfo {volume} {107}},\ \bibinfo {pages} {365} (\bibinfo {year}
  {1976})}\BibitemShut {NoStop}%
\bibitem [{\citenamefont {Wu}\ and\ \citenamefont {Yang}(1977)}]{Wu77}%
  \BibitemOpen
  \bibfield  {author} {\bibinfo {author} {\bibfnamefont {T.~T.}\ \bibnamefont
  {Wu}}\ and\ \bibinfo {author} {\bibfnamefont {C.~N.}\ \bibnamefont {Yang}},\
  }\bibfield  {title} {\bibinfo {title} {Some properties of monopole
  harmonics},\ }\href {https://doi.org/10.1103/PhysRevD.16.1018} {\bibfield
  {journal} {\bibinfo  {journal} {Phys. Rev. D}\ }\textbf {\bibinfo {volume}
  {16}},\ \bibinfo {pages} {1018} (\bibinfo {year} {1977})}\BibitemShut
  {NoStop}%
\bibitem [{\citenamefont {Greiter}(2011)}]{Greiter11}%
  \BibitemOpen
  \bibfield  {author} {\bibinfo {author} {\bibfnamefont {M.}~\bibnamefont
  {Greiter}},\ }\bibfield  {title} {\bibinfo {title} {{Landau} level
  quantization on the sphere},\ }\href
  {https://doi.org/10.1103/PhysRevB.83.115129} {\bibfield  {journal} {\bibinfo
  {journal} {Phys. Rev. B}\ }\textbf {\bibinfo {volume} {83}},\ \bibinfo
  {pages} {115129} (\bibinfo {year} {2011})}\BibitemShut {NoStop}%
\bibitem [{\citenamefont {Balram}\ \emph
  {et~al.}(2013{\natexlab{b}})\citenamefont {Balram}, \citenamefont {W\'ojs},\
  and\ \citenamefont {Jain}}]{Balram13}%
  \BibitemOpen
  \bibfield  {author} {\bibinfo {author} {\bibfnamefont {A.~C.}\ \bibnamefont
  {Balram}}, \bibinfo {author} {\bibfnamefont {A.}~\bibnamefont {W\'ojs}},\
  and\ \bibinfo {author} {\bibfnamefont {J.~K.}\ \bibnamefont {Jain}},\
  }\bibfield  {title} {\bibinfo {title} {State counting for excited bands of
  the fractional quantum {Hall} effect: Exclusion rules for bound excitons},\
  }\href {https://doi.org/10.1103/PhysRevB.88.205312} {\bibfield  {journal}
  {\bibinfo  {journal} {Phys. Rev. B}\ }\textbf {\bibinfo {volume} {88}},\
  \bibinfo {pages} {205312} (\bibinfo {year} {2013}{\natexlab{b}})}\BibitemShut
  {NoStop}%
\bibitem [{\citenamefont {Johri}\ \emph {et~al.}(2014)\citenamefont {Johri},
  \citenamefont {Papi\ifmmode~\acute{c}\else \'{c}\fi{}}, \citenamefont
  {Bhatt},\ and\ \citenamefont {Schmitteckert}}]{Johri14}%
  \BibitemOpen
  \bibfield  {author} {\bibinfo {author} {\bibfnamefont {S.}~\bibnamefont
  {Johri}}, \bibinfo {author} {\bibfnamefont {Z.}~\bibnamefont
  {Papi\ifmmode~\acute{c}\else \'{c}\fi{}}}, \bibinfo {author} {\bibfnamefont
  {R.~N.}\ \bibnamefont {Bhatt}},\ and\ \bibinfo {author} {\bibfnamefont
  {P.}~\bibnamefont {Schmitteckert}},\ }\bibfield  {title} {\bibinfo {title}
  {Quasiholes of $\frac{1}{3}$ and $\frac{7}{3}$ quantum {Hall} states: Size
  estimates via exact diagonalization and density-matrix renormalization
  group},\ }\href {https://doi.org/10.1103/PhysRevB.89.115124} {\bibfield
  {journal} {\bibinfo  {journal} {Phys. Rev. B}\ }\textbf {\bibinfo {volume}
  {89}},\ \bibinfo {pages} {115124} (\bibinfo {year} {2014})}\BibitemShut
  {NoStop}%
\bibitem [{\citenamefont {Ji}\ \emph {et~al.}(2024)\citenamefont {Ji},
  \citenamefont {Bose}, \citenamefont {Balram},\ and\ \citenamefont
  {Yang}}]{Ji24}%
  \BibitemOpen
  \bibfield  {author} {\bibinfo {author} {\bibfnamefont {G.}~\bibnamefont
  {Ji}}, \bibinfo {author} {\bibfnamefont {K.}~\bibnamefont {Bose}}, \bibinfo
  {author} {\bibfnamefont {A.~C.}\ \bibnamefont {Balram}},\ and\ \bibinfo
  {author} {\bibfnamefont {B.}~\bibnamefont {Yang}},\ }\href@noop {} {\bibinfo
  {title} {Universal modelling of emergent oscillations in fractional quantum
  hall fluids}} (\bibinfo {year} {2024}),\ \Eprint
  {https://arxiv.org/abs/2401.06856} {arXiv:2401.06856 [cond-mat.str-el]}
  \BibitemShut {NoStop}%
\bibitem [{\citenamefont {Balram}\ \emph {et~al.}(2020)\citenamefont {Balram},
  \citenamefont {Jain},\ and\ \citenamefont {Barkeshli}}]{Balram20}%
  \BibitemOpen
  \bibfield  {author} {\bibinfo {author} {\bibfnamefont {A.~C.}\ \bibnamefont
  {Balram}}, \bibinfo {author} {\bibfnamefont {J.~K.}\ \bibnamefont {Jain}},\
  and\ \bibinfo {author} {\bibfnamefont {M.}~\bibnamefont {Barkeshli}},\
  }\bibfield  {title} {\bibinfo {title} {${\mathbb{z}}_{n}$ superconductivity
  of composite bosons and the $7/3$ fractional quantum {Hall} effect},\ }\href
  {https://doi.org/10.1103/PhysRevResearch.2.013349} {\bibfield  {journal}
  {\bibinfo  {journal} {Phys. Rev. Research}\ }\textbf {\bibinfo {volume}
  {2}},\ \bibinfo {pages} {013349} (\bibinfo {year} {2020})}\BibitemShut
  {NoStop}%
\bibitem [{\citenamefont {Moore}\ and\ \citenamefont {Read}(1991)}]{Moore91}%
  \BibitemOpen
  \bibfield  {author} {\bibinfo {author} {\bibfnamefont {G.}~\bibnamefont
  {Moore}}\ and\ \bibinfo {author} {\bibfnamefont {N.}~\bibnamefont {Read}},\
  }\bibfield  {title} {\bibinfo {title} {Nonabelions in the fractional quantum
  {Hall} effect},\ }\href {https://doi.org/10.1016/0550-3213(91)90407-O}
  {\bibfield  {journal} {\bibinfo  {journal} {Nucl. Phys. B}\ }\textbf
  {\bibinfo {volume} {360}},\ \bibinfo {pages} {362 } (\bibinfo {year}
  {1991})}\BibitemShut {NoStop}%
\bibitem [{\citenamefont {Wilczek}\ and\ \citenamefont
  {Zee}(1983)}]{Wilczek83}%
  \BibitemOpen
  \bibfield  {author} {\bibinfo {author} {\bibfnamefont {F.}~\bibnamefont
  {Wilczek}}\ and\ \bibinfo {author} {\bibfnamefont {A.}~\bibnamefont {Zee}},\
  }\bibfield  {title} {\bibinfo {title} {Linking numbers, spin, and statistics
  of solitons},\ }\href {https://doi.org/10.1103/PhysRevLett.51.2250}
  {\bibfield  {journal} {\bibinfo  {journal} {Phys. Rev. Lett.}\ }\textbf
  {\bibinfo {volume} {51}},\ \bibinfo {pages} {2250} (\bibinfo {year}
  {1983})}\BibitemShut {NoStop}%
\bibitem [{\citenamefont {Binder}\ and\ \citenamefont
  {Heermann}(2010)}]{Binder10}%
  \BibitemOpen
  \bibfield  {author} {\bibinfo {author} {\bibfnamefont {K.}~\bibnamefont
  {Binder}}\ and\ \bibinfo {author} {\bibfnamefont {D.}~\bibnamefont
  {Heermann}},\ }\href@noop {} {\emph {\bibinfo {title} {{Monte} {Carlo}
  Simulation in Statistical Physics}}}\ (\bibinfo  {publisher} {Springer-Verlag
  Berlin Heidelberg},\ \bibinfo {year} {2010})\BibitemShut {NoStop}%
\bibitem [{\citenamefont {Macaluso}\ \emph {et~al.}(2019)\citenamefont
  {Macaluso}, \citenamefont {Comparin}, \citenamefont {Mazza},\ and\
  \citenamefont {Carusotto}}]{Macaluso19}%
  \BibitemOpen
  \bibfield  {author} {\bibinfo {author} {\bibfnamefont {E.}~\bibnamefont
  {Macaluso}}, \bibinfo {author} {\bibfnamefont {T.}~\bibnamefont {Comparin}},
  \bibinfo {author} {\bibfnamefont {L.}~\bibnamefont {Mazza}},\ and\ \bibinfo
  {author} {\bibfnamefont {I.}~\bibnamefont {Carusotto}},\ }\bibfield  {title}
  {\bibinfo {title} {Fusion channels of non-{Abelian} anyons from
  angular-momentum and density-profile measurements},\ }\href
  {https://doi.org/10.1103/PhysRevLett.123.266801} {\bibfield  {journal}
  {\bibinfo  {journal} {Phys. Rev. Lett.}\ }\textbf {\bibinfo {volume} {123}},\
  \bibinfo {pages} {266801} (\bibinfo {year} {2019})}\BibitemShut {NoStop}%
\bibitem [{\citenamefont {Wen}(1995)}]{Wen95}%
  \BibitemOpen
  \bibfield  {author} {\bibinfo {author} {\bibfnamefont {X.-G.}\ \bibnamefont
  {Wen}},\ }\bibfield  {title} {\bibinfo {title} {Topological orders and edge
  excitations in fractional quantum {Hall} states},\ }\href
  {https://doi.org/10.1080/00018739500101566} {\bibfield  {journal} {\bibinfo
  {journal} {Advances in Physics}\ }\textbf {\bibinfo {volume} {44}},\ \bibinfo
  {pages} {405} (\bibinfo {year} {1995})},\ \Eprint
  {https://arxiv.org/abs/http://www.tandfonline.com/doi/pdf/10.1080/00018739500101566}
  {http://www.tandfonline.com/doi/pdf/10.1080/00018739500101566} \BibitemShut
  {NoStop}%
\bibitem [{\citenamefont {Gattu}\ \emph {et~al.}(2024)\citenamefont {Gattu},
  \citenamefont {Sreejith},\ and\ \citenamefont {Jain}}]{Gattu23}%
  \BibitemOpen
  \bibfield  {author} {\bibinfo {author} {\bibfnamefont {M.}~\bibnamefont
  {Gattu}}, \bibinfo {author} {\bibfnamefont {G.~J.}\ \bibnamefont
  {Sreejith}},\ and\ \bibinfo {author} {\bibfnamefont {J.~K.}\ \bibnamefont
  {Jain}},\ }\bibfield  {title} {\bibinfo {title} {Scanning tunneling
  microscopy of fractional quantum {Hall} states: Spectroscopy of
  composite-fermion bound states},\ }\href
  {https://doi.org/10.1103/PhysRevB.109.L201123} {\bibfield  {journal}
  {\bibinfo  {journal} {Phys. Rev. B}\ }\textbf {\bibinfo {volume} {109}},\
  \bibinfo {pages} {L201123} (\bibinfo {year} {2024})}\BibitemShut {NoStop}%
\bibitem [{\citenamefont {Willett}\ \emph {et~al.}(1987)\citenamefont
  {Willett}, \citenamefont {Eisenstein}, \citenamefont {St\"ormer},
  \citenamefont {Tsui}, \citenamefont {Gossard},\ and\ \citenamefont
  {English}}]{Willett87}%
  \BibitemOpen
  \bibfield  {author} {\bibinfo {author} {\bibfnamefont {R.}~\bibnamefont
  {Willett}}, \bibinfo {author} {\bibfnamefont {J.~P.}\ \bibnamefont
  {Eisenstein}}, \bibinfo {author} {\bibfnamefont {H.~L.}\ \bibnamefont
  {St\"ormer}}, \bibinfo {author} {\bibfnamefont {D.~C.}\ \bibnamefont {Tsui}},
  \bibinfo {author} {\bibfnamefont {A.~C.}\ \bibnamefont {Gossard}},\ and\
  \bibinfo {author} {\bibfnamefont {J.~H.}\ \bibnamefont {English}},\
  }\bibfield  {title} {\bibinfo {title} {Observation of an even-denominator
  quantum number in the fractional quantum {Hall} effect},\ }\href
  {https://doi.org/10.1103/PhysRevLett.59.1776} {\bibfield  {journal} {\bibinfo
   {journal} {Phys. Rev. Lett.}\ }\textbf {\bibinfo {volume} {59}},\ \bibinfo
  {pages} {1776} (\bibinfo {year} {1987})}\BibitemShut {NoStop}%
\bibitem [{\citenamefont {Xia}\ \emph {et~al.}(2004)\citenamefont {Xia},
  \citenamefont {Pan}, \citenamefont {Vicente}, \citenamefont {Adams},
  \citenamefont {Sullivan}, \citenamefont {Stormer}, \citenamefont {Tsui},
  \citenamefont {Pfeiffer}, \citenamefont {Baldwin},\ and\ \citenamefont
  {West}}]{Xia04}%
  \BibitemOpen
  \bibfield  {author} {\bibinfo {author} {\bibfnamefont {J.~S.}\ \bibnamefont
  {Xia}}, \bibinfo {author} {\bibfnamefont {W.}~\bibnamefont {Pan}}, \bibinfo
  {author} {\bibfnamefont {C.~L.}\ \bibnamefont {Vicente}}, \bibinfo {author}
  {\bibfnamefont {E.~D.}\ \bibnamefont {Adams}}, \bibinfo {author}
  {\bibfnamefont {N.~S.}\ \bibnamefont {Sullivan}}, \bibinfo {author}
  {\bibfnamefont {H.~L.}\ \bibnamefont {Stormer}}, \bibinfo {author}
  {\bibfnamefont {D.~C.}\ \bibnamefont {Tsui}}, \bibinfo {author}
  {\bibfnamefont {L.~N.}\ \bibnamefont {Pfeiffer}}, \bibinfo {author}
  {\bibfnamefont {K.~W.}\ \bibnamefont {Baldwin}},\ and\ \bibinfo {author}
  {\bibfnamefont {K.~W.}\ \bibnamefont {West}},\ }\bibfield  {title} {\bibinfo
  {title} {Electron correlation in the second {Landau} level: A competition
  between many nearly degenerate quantum phases},\ }\href
  {https://doi.org/10.1103/PhysRevLett.93.176809} {\bibfield  {journal}
  {\bibinfo  {journal} {Phys. Rev. Lett.}\ }\textbf {\bibinfo {volume} {93}},\
  \bibinfo {pages} {176809} (\bibinfo {year} {2004})}\BibitemShut {NoStop}%
\bibitem [{\citenamefont {Kumar}\ \emph {et~al.}(2010)\citenamefont {Kumar},
  \citenamefont {Cs\'athy}, \citenamefont {Manfra}, \citenamefont {Pfeiffer},\
  and\ \citenamefont {West}}]{Kumar10}%
  \BibitemOpen
  \bibfield  {author} {\bibinfo {author} {\bibfnamefont {A.}~\bibnamefont
  {Kumar}}, \bibinfo {author} {\bibfnamefont {G.~A.}\ \bibnamefont {Cs\'athy}},
  \bibinfo {author} {\bibfnamefont {M.~J.}\ \bibnamefont {Manfra}}, \bibinfo
  {author} {\bibfnamefont {L.~N.}\ \bibnamefont {Pfeiffer}},\ and\ \bibinfo
  {author} {\bibfnamefont {K.~W.}\ \bibnamefont {West}},\ }\bibfield  {title}
  {\bibinfo {title} {Nonconventional odd-denominator fractional quantum {Hall}
  states in the second {Landau} level},\ }\href
  {https://doi.org/10.1103/PhysRevLett.105.246808} {\bibfield  {journal}
  {\bibinfo  {journal} {Phys. Rev. Lett.}\ }\textbf {\bibinfo {volume} {105}},\
  \bibinfo {pages} {246808} (\bibinfo {year} {2010})}\BibitemShut {NoStop}%
\bibitem [{\citenamefont {Pan}\ \emph {et~al.}(2003)\citenamefont {Pan},
  \citenamefont {Stormer}, \citenamefont {Tsui}, \citenamefont {Pfeiffer},
  \citenamefont {Baldwin},\ and\ \citenamefont {West}}]{Pan03}%
  \BibitemOpen
  \bibfield  {author} {\bibinfo {author} {\bibfnamefont {W.}~\bibnamefont
  {Pan}}, \bibinfo {author} {\bibfnamefont {H.~L.}\ \bibnamefont {Stormer}},
  \bibinfo {author} {\bibfnamefont {D.~C.}\ \bibnamefont {Tsui}}, \bibinfo
  {author} {\bibfnamefont {L.~N.}\ \bibnamefont {Pfeiffer}}, \bibinfo {author}
  {\bibfnamefont {K.~W.}\ \bibnamefont {Baldwin}},\ and\ \bibinfo {author}
  {\bibfnamefont {K.~W.}\ \bibnamefont {West}},\ }\bibfield  {title} {\bibinfo
  {title} {Fractional quantum {Hall} effect of composite fermions},\ }\href
  {https://doi.org/10.1103/PhysRevLett.90.016801} {\bibfield  {journal}
  {\bibinfo  {journal} {Phys. Rev. Lett.}\ }\textbf {\bibinfo {volume} {90}},\
  \bibinfo {pages} {016801} (\bibinfo {year} {2003})}\BibitemShut {NoStop}%
\bibitem [{\citenamefont {Samkharadze}\ \emph {et~al.}(2015)\citenamefont
  {Samkharadze}, \citenamefont {Arnold}, \citenamefont {Pfeiffer},
  \citenamefont {West},\ and\ \citenamefont {Cs\'athy}}]{Samkharadze15b}%
  \BibitemOpen
  \bibfield  {author} {\bibinfo {author} {\bibfnamefont {N.}~\bibnamefont
  {Samkharadze}}, \bibinfo {author} {\bibfnamefont {I.}~\bibnamefont {Arnold}},
  \bibinfo {author} {\bibfnamefont {L.~N.}\ \bibnamefont {Pfeiffer}}, \bibinfo
  {author} {\bibfnamefont {K.~W.}\ \bibnamefont {West}},\ and\ \bibinfo
  {author} {\bibfnamefont {G.~A.}\ \bibnamefont {Cs\'athy}},\ }\bibfield
  {title} {\bibinfo {title} {Observation of incompressibility at $\nu=4/11$ and
  $\nu=5/13$},\ }\href {https://doi.org/10.1103/PhysRevB.91.081109} {\bibfield
  {journal} {\bibinfo  {journal} {Phys. Rev. B}\ }\textbf {\bibinfo {volume}
  {91}},\ \bibinfo {pages} {081109} (\bibinfo {year} {2015})}\BibitemShut
  {NoStop}%
\bibitem [{\citenamefont {Pan}\ \emph {et~al.}(2015)\citenamefont {Pan},
  \citenamefont {Baldwin}, \citenamefont {West}, \citenamefont {Pfeiffer},\
  and\ \citenamefont {Tsui}}]{Pan15}%
  \BibitemOpen
  \bibfield  {author} {\bibinfo {author} {\bibfnamefont {W.}~\bibnamefont
  {Pan}}, \bibinfo {author} {\bibfnamefont {K.~W.}\ \bibnamefont {Baldwin}},
  \bibinfo {author} {\bibfnamefont {K.~W.}\ \bibnamefont {West}}, \bibinfo
  {author} {\bibfnamefont {L.~N.}\ \bibnamefont {Pfeiffer}},\ and\ \bibinfo
  {author} {\bibfnamefont {D.~C.}\ \bibnamefont {Tsui}},\ }\bibfield  {title}
  {\bibinfo {title} {Fractional quantum {Hall} effect at {Landau} level filling
  $\ensuremath{\nu}=4/11$},\ }\href
  {https://doi.org/10.1103/PhysRevB.91.041301} {\bibfield  {journal} {\bibinfo
  {journal} {Phys. Rev. B}\ }\textbf {\bibinfo {volume} {91}},\ \bibinfo
  {pages} {041301} (\bibinfo {year} {2015})}\BibitemShut {NoStop}%
\bibitem [{\citenamefont {Kumar}\ \emph {et~al.}(2019)\citenamefont {Kumar},
  \citenamefont {Raghu},\ and\ \citenamefont {Mulligan}}]{Kumar18}%
  \BibitemOpen
  \bibfield  {author} {\bibinfo {author} {\bibfnamefont {P.}~\bibnamefont
  {Kumar}}, \bibinfo {author} {\bibfnamefont {S.}~\bibnamefont {Raghu}},\ and\
  \bibinfo {author} {\bibfnamefont {M.}~\bibnamefont {Mulligan}},\ }\bibfield
  {title} {\bibinfo {title} {Composite fermion {Hall} conductivity and the
  half-filled {Landau} level},\ }\href
  {https://doi.org/10.1103/PhysRevB.99.235114} {\bibfield  {journal} {\bibinfo
  {journal} {Phys. Rev. B}\ }\textbf {\bibinfo {volume} {99}},\ \bibinfo
  {pages} {235114} (\bibinfo {year} {2019})}\BibitemShut {NoStop}%
\bibitem [{\citenamefont {Sitko}\ \emph {et~al.}(1996)\citenamefont {Sitko},
  \citenamefont {Yi}, \citenamefont {Yi},\ and\ \citenamefont
  {Quinn}}]{Sitko96}%
  \BibitemOpen
  \bibfield  {author} {\bibinfo {author} {\bibfnamefont {P.}~\bibnamefont
  {Sitko}}, \bibinfo {author} {\bibfnamefont {S.~N.}\ \bibnamefont {Yi}},
  \bibinfo {author} {\bibfnamefont {K.~S.}\ \bibnamefont {Yi}},\ and\ \bibinfo
  {author} {\bibfnamefont {J.~J.}\ \bibnamefont {Quinn}},\ }\bibfield  {title}
  {\bibinfo {title} {"fermi liquid" shell model approach to composite fermion
  excitation spectra in fractional quantum {Hall} states},\ }\href
  {https://doi.org/10.1103/PhysRevLett.76.3396} {\bibfield  {journal} {\bibinfo
   {journal} {Phys. Rev. Lett.}\ }\textbf {\bibinfo {volume} {76}},\ \bibinfo
  {pages} {3396} (\bibinfo {year} {1996})}\BibitemShut {NoStop}%
\bibitem [{\citenamefont {Read}\ and\ \citenamefont {Green}(2000)}]{Read00}%
  \BibitemOpen
  \bibfield  {author} {\bibinfo {author} {\bibfnamefont {N.}~\bibnamefont
  {Read}}\ and\ \bibinfo {author} {\bibfnamefont {D.}~\bibnamefont {Green}},\
  }\bibfield  {title} {\bibinfo {title} {Paired states of fermions in two
  dimensions with breaking of parity and time-reversal symmetries and the
  fractional quantum {Hall} effect},\ }\href
  {https://doi.org/10.1103/PhysRevB.61.10267} {\bibfield  {journal} {\bibinfo
  {journal} {Phys. Rev. B}\ }\textbf {\bibinfo {volume} {61}},\ \bibinfo
  {pages} {10267} (\bibinfo {year} {2000})}\BibitemShut {NoStop}%
\bibitem [{\citenamefont {Scarola}\ \emph {et~al.}(2002)\citenamefont
  {Scarola}, \citenamefont {Lee},\ and\ \citenamefont {Jain}}]{Scarola02}%
  \BibitemOpen
  \bibfield  {author} {\bibinfo {author} {\bibfnamefont {V.~W.}\ \bibnamefont
  {Scarola}}, \bibinfo {author} {\bibfnamefont {S.-Y.}\ \bibnamefont {Lee}},\
  and\ \bibinfo {author} {\bibfnamefont {J.~K.}\ \bibnamefont {Jain}},\
  }\bibfield  {title} {\bibinfo {title} {Excitation gaps of incompressible
  composite fermion states: Approach to the {Fermi} sea},\ }\href
  {https://doi.org/10.1103/PhysRevB.66.155320} {\bibfield  {journal} {\bibinfo
  {journal} {Phys. Rev. B}\ }\textbf {\bibinfo {volume} {66}},\ \bibinfo
  {pages} {155320} (\bibinfo {year} {2002})}\BibitemShut {NoStop}%
\bibitem [{\citenamefont {Mukherjee}\ \emph {et~al.}(2014)\citenamefont
  {Mukherjee}, \citenamefont {Mandal}, \citenamefont {Wu}, \citenamefont
  {W\'ojs},\ and\ \citenamefont {Jain}}]{Mukherjee14}%
  \BibitemOpen
  \bibfield  {author} {\bibinfo {author} {\bibfnamefont {S.}~\bibnamefont
  {Mukherjee}}, \bibinfo {author} {\bibfnamefont {S.~S.}\ \bibnamefont
  {Mandal}}, \bibinfo {author} {\bibfnamefont {Y.-H.}\ \bibnamefont {Wu}},
  \bibinfo {author} {\bibfnamefont {A.}~\bibnamefont {W\'ojs}},\ and\ \bibinfo
  {author} {\bibfnamefont {J.~K.}\ \bibnamefont {Jain}},\ }\bibfield  {title}
  {\bibinfo {title} {Enigmatic $4/11$ state: A prototype for unconventional
  fractional quantum {Hall} effect},\ }\href
  {https://doi.org/10.1103/PhysRevLett.112.016801} {\bibfield  {journal}
  {\bibinfo  {journal} {Phys. Rev. Lett.}\ }\textbf {\bibinfo {volume} {112}},\
  \bibinfo {pages} {016801} (\bibinfo {year} {2014})}\BibitemShut {NoStop}%
\bibitem [{\citenamefont {Balram}\ \emph {et~al.}(2015)\citenamefont {Balram},
  \citenamefont {T\"oke}, \citenamefont {W\'ojs},\ and\ \citenamefont
  {Jain}}]{Balram15}%
  \BibitemOpen
  \bibfield  {author} {\bibinfo {author} {\bibfnamefont {A.~C.}\ \bibnamefont
  {Balram}}, \bibinfo {author} {\bibfnamefont {C.}~\bibnamefont {T\"oke}},
  \bibinfo {author} {\bibfnamefont {A.}~\bibnamefont {W\'ojs}},\ and\ \bibinfo
  {author} {\bibfnamefont {J.~K.}\ \bibnamefont {Jain}},\ }\bibfield  {title}
  {\bibinfo {title} {Phase diagram of fractional quantum {Hall} effect of
  composite fermions in multicomponent systems},\ }\href
  {https://doi.org/10.1103/PhysRevB.91.045109} {\bibfield  {journal} {\bibinfo
  {journal} {Phys. Rev. B}\ }\textbf {\bibinfo {volume} {91}},\ \bibinfo
  {pages} {045109} (\bibinfo {year} {2015})}\BibitemShut {NoStop}%
\bibitem [{\citenamefont {Balram}(2016)}]{Balram16c}%
  \BibitemOpen
  \bibfield  {author} {\bibinfo {author} {\bibfnamefont {A.~C.}\ \bibnamefont
  {Balram}},\ }\bibfield  {title} {\bibinfo {title} {Interacting composite
  fermions: Nature of the 4/5, 5/7, 6/7, and 6/17 fractional quantum {Hall}
  states},\ }\href {https://doi.org/10.1103/PhysRevB.94.165303} {\bibfield
  {journal} {\bibinfo  {journal} {Phys. Rev. B}\ }\textbf {\bibinfo {volume}
  {94}},\ \bibinfo {pages} {165303} (\bibinfo {year} {2016})}\BibitemShut
  {NoStop}%
\bibitem [{\citenamefont {Jain}(1989{\natexlab{b}})}]{Jain89b}%
  \BibitemOpen
  \bibfield  {author} {\bibinfo {author} {\bibfnamefont {J.~K.}\ \bibnamefont
  {Jain}},\ }\bibfield  {title} {\bibinfo {title} {Incompressible quantum
  {Hall} states},\ }\href {https://doi.org/10.1103/PhysRevB.40.8079} {\bibfield
   {journal} {\bibinfo  {journal} {Phys. Rev. B}\ }\textbf {\bibinfo {volume}
  {40}},\ \bibinfo {pages} {8079} (\bibinfo {year}
  {1989}{\natexlab{b}})}\BibitemShut {NoStop}%
\bibitem [{\citenamefont {Wu}\ \emph {et~al.}(2017)\citenamefont {Wu},
  \citenamefont {Shi},\ and\ \citenamefont {Jain}}]{Wu17}%
  \BibitemOpen
  \bibfield  {author} {\bibinfo {author} {\bibfnamefont {Y.}~\bibnamefont
  {Wu}}, \bibinfo {author} {\bibfnamefont {T.}~\bibnamefont {Shi}},\ and\
  \bibinfo {author} {\bibfnamefont {J.~K.}\ \bibnamefont {Jain}},\ }\bibfield
  {title} {\bibinfo {title} {Non-abelian parton fractional quantum {Hall}
  effect in multilayer graphene},\ }\href
  {https://doi.org/10.1021/acs.nanolett.7b01080} {\bibfield  {journal}
  {\bibinfo  {journal} {Nano Letters}\ }\textbf {\bibinfo {volume} {17}},\
  \bibinfo {pages} {4643} (\bibinfo {year} {2017})},\ \bibinfo {note} {pMID:
  28649831},\ \Eprint
  {https://arxiv.org/abs/http://dx.doi.org/10.1021/acs.nanolett.7b01080}
  {http://dx.doi.org/10.1021/acs.nanolett.7b01080} \BibitemShut {NoStop}%
\bibitem [{\citenamefont {Balram}\ \emph {et~al.}(2018)\citenamefont {Balram},
  \citenamefont {Barkeshli},\ and\ \citenamefont {Rudner}}]{Balram18}%
  \BibitemOpen
  \bibfield  {author} {\bibinfo {author} {\bibfnamefont {A.~C.}\ \bibnamefont
  {Balram}}, \bibinfo {author} {\bibfnamefont {M.}~\bibnamefont {Barkeshli}},\
  and\ \bibinfo {author} {\bibfnamefont {M.~S.}\ \bibnamefont {Rudner}},\
  }\bibfield  {title} {\bibinfo {title} {Parton construction of a wave function
  in the anti-{Pfaffian} phase},\ }\href
  {https://doi.org/10.1103/PhysRevB.98.035127} {\bibfield  {journal} {\bibinfo
  {journal} {Phys. Rev. B}\ }\textbf {\bibinfo {volume} {98}},\ \bibinfo
  {pages} {035127} (\bibinfo {year} {2018})}\BibitemShut {NoStop}%
\bibitem [{\citenamefont {Balram}\ \emph {et~al.}(2019)\citenamefont {Balram},
  \citenamefont {Barkeshli},\ and\ \citenamefont {Rudner}}]{Balram19}%
  \BibitemOpen
  \bibfield  {author} {\bibinfo {author} {\bibfnamefont {A.~C.}\ \bibnamefont
  {Balram}}, \bibinfo {author} {\bibfnamefont {M.}~\bibnamefont {Barkeshli}},\
  and\ \bibinfo {author} {\bibfnamefont {M.~S.}\ \bibnamefont {Rudner}},\
  }\bibfield  {title} {\bibinfo {title} {Parton construction of
  particle-hole-conjugate {Read}-{Rezayi} parafermion fractional quantum {Hall}
  states and beyond},\ }\href {https://doi.org/10.1103/PhysRevB.99.241108}
  {\bibfield  {journal} {\bibinfo  {journal} {Phys. Rev. B}\ }\textbf {\bibinfo
  {volume} {99}},\ \bibinfo {pages} {241108} (\bibinfo {year}
  {2019})}\BibitemShut {NoStop}%
\bibitem [{\citenamefont {Faugno}\ \emph {et~al.}(2019)\citenamefont {Faugno},
  \citenamefont {Balram}, \citenamefont {Barkeshli},\ and\ \citenamefont
  {Jain}}]{Faugno19}%
  \BibitemOpen
  \bibfield  {author} {\bibinfo {author} {\bibfnamefont {W.~N.}\ \bibnamefont
  {Faugno}}, \bibinfo {author} {\bibfnamefont {A.~C.}\ \bibnamefont {Balram}},
  \bibinfo {author} {\bibfnamefont {M.}~\bibnamefont {Barkeshli}},\ and\
  \bibinfo {author} {\bibfnamefont {J.~K.}\ \bibnamefont {Jain}},\ }\bibfield
  {title} {\bibinfo {title} {Prediction of a non-{Abelian} fractional quantum
  {Hall} state with $f$-wave pairing of composite fermions in wide quantum
  wells},\ }\href {https://doi.org/10.1103/PhysRevLett.123.016802} {\bibfield
  {journal} {\bibinfo  {journal} {Phys. Rev. Lett.}\ }\textbf {\bibinfo
  {volume} {123}},\ \bibinfo {pages} {016802} (\bibinfo {year}
  {2019})}\BibitemShut {NoStop}%
\bibitem [{\citenamefont {Faugno}\ \emph {et~al.}(2020)\citenamefont {Faugno},
  \citenamefont {Jain},\ and\ \citenamefont {Balram}}]{Faugno20a}%
  \BibitemOpen
  \bibfield  {author} {\bibinfo {author} {\bibfnamefont {W.~N.}\ \bibnamefont
  {Faugno}}, \bibinfo {author} {\bibfnamefont {J.~K.}\ \bibnamefont {Jain}},\
  and\ \bibinfo {author} {\bibfnamefont {A.~C.}\ \bibnamefont {Balram}},\
  }\bibfield  {title} {\bibinfo {title} {Non-abelian fractional quantum {Hall}
  state at $3/7$-filled {Landau} level},\ }\href
  {https://doi.org/10.1103/PhysRevResearch.2.033223} {\bibfield  {journal}
  {\bibinfo  {journal} {Phys. Rev. Research}\ }\textbf {\bibinfo {volume}
  {2}},\ \bibinfo {pages} {033223} (\bibinfo {year} {2020})}\BibitemShut
  {NoStop}%
\bibitem [{\citenamefont {Balram}\ and\ \citenamefont
  {W\'ojs}(2020)}]{Balram20b}%
  \BibitemOpen
  \bibfield  {author} {\bibinfo {author} {\bibfnamefont {A.~C.}\ \bibnamefont
  {Balram}}\ and\ \bibinfo {author} {\bibfnamefont {A.}~\bibnamefont
  {W\'ojs}},\ }\bibfield  {title} {\bibinfo {title} {Fractional quantum {Hall}
  effect at $\ensuremath{\nu}=2+4/9$},\ }\href
  {https://doi.org/10.1103/PhysRevResearch.2.032035} {\bibfield  {journal}
  {\bibinfo  {journal} {Phys. Rev. Research}\ }\textbf {\bibinfo {volume}
  {2}},\ \bibinfo {pages} {032035} (\bibinfo {year} {2020})}\BibitemShut
  {NoStop}%
\bibitem [{\citenamefont {Faugno}\ \emph {et~al.}(2021)\citenamefont {Faugno},
  \citenamefont {Zhao}, \citenamefont {Balram}, \citenamefont {Jolicoeur},\
  and\ \citenamefont {Jain}}]{Faugno21}%
  \BibitemOpen
  \bibfield  {author} {\bibinfo {author} {\bibfnamefont {W.~N.}\ \bibnamefont
  {Faugno}}, \bibinfo {author} {\bibfnamefont {T.}~\bibnamefont {Zhao}},
  \bibinfo {author} {\bibfnamefont {A.~C.}\ \bibnamefont {Balram}}, \bibinfo
  {author} {\bibfnamefont {T.}~\bibnamefont {Jolicoeur}},\ and\ \bibinfo
  {author} {\bibfnamefont {J.~K.}\ \bibnamefont {Jain}},\ }\bibfield  {title}
  {\bibinfo {title} {Unconventional ${\mathbb{z}}_{n}$ parton states at
  $\ensuremath{\nu}=7/3$: {Role} of finite width},\ }\href
  {https://doi.org/10.1103/PhysRevB.103.085303} {\bibfield  {journal} {\bibinfo
   {journal} {Phys. Rev. B}\ }\textbf {\bibinfo {volume} {103}},\ \bibinfo
  {pages} {085303} (\bibinfo {year} {2021})}\BibitemShut {NoStop}%
\bibitem [{\citenamefont {Balram}(2021{\natexlab{a}})}]{Balram21}%
  \BibitemOpen
  \bibfield  {author} {\bibinfo {author} {\bibfnamefont {A.~C.}\ \bibnamefont
  {Balram}},\ }\bibfield  {title} {\bibinfo {title} {{A non-{Abelian} parton
  state for the $\ensuremath{\nu}=2+3/8$ fractional quantum {Hall} effect}},\
  }\href {https://doi.org/10.21468/SciPostPhys.10.4.083} {\bibfield  {journal}
  {\bibinfo  {journal} {SciPost Phys.}\ }\textbf {\bibinfo {volume} {10}},\
  \bibinfo {pages} {83} (\bibinfo {year} {2021}{\natexlab{a}})}\BibitemShut
  {NoStop}%
\bibitem [{\citenamefont {Balram}\ and\ \citenamefont
  {W\'ojs}(2021)}]{Balram21a}%
  \BibitemOpen
  \bibfield  {author} {\bibinfo {author} {\bibfnamefont {A.~C.}\ \bibnamefont
  {Balram}}\ and\ \bibinfo {author} {\bibfnamefont {A.}~\bibnamefont
  {W\'ojs}},\ }\bibfield  {title} {\bibinfo {title} {Parton wave function for
  the fractional quantum hall effect at $\ensuremath{\nu}=6/17$},\ }\href
  {https://doi.org/10.1103/PhysRevResearch.3.033087} {\bibfield  {journal}
  {\bibinfo  {journal} {Phys. Rev. Research}\ }\textbf {\bibinfo {volume}
  {3}},\ \bibinfo {pages} {033087} (\bibinfo {year} {2021})}\BibitemShut
  {NoStop}%
\bibitem [{\citenamefont {Balram}(2022)}]{Balram21b}%
  \BibitemOpen
  \bibfield  {author} {\bibinfo {author} {\bibfnamefont {A.~C.}\ \bibnamefont
  {Balram}},\ }\bibfield  {title} {\bibinfo {title} {Transitions from {Abelian}
  composite fermion to non-{Abelian} parton fractional quantum {Hall} states in
  the zeroth {Landau} level of bilayer graphene},\ }\href
  {https://doi.org/10.1103/PhysRevB.105.L121406} {\bibfield  {journal}
  {\bibinfo  {journal} {Phys. Rev. B}\ }\textbf {\bibinfo {volume} {105}},\
  \bibinfo {pages} {L121406} (\bibinfo {year} {2022})}\BibitemShut {NoStop}%
\bibitem [{\citenamefont {Balram}(2021{\natexlab{b}})}]{Balram21c}%
  \BibitemOpen
  \bibfield  {author} {\bibinfo {author} {\bibfnamefont {A.~C.}\ \bibnamefont
  {Balram}},\ }\bibfield  {title} {\bibinfo {title} {Abelian parton state for
  the $\ensuremath{\nu}=4/11$ fractional quantum {Hall} effect},\ }\href
  {https://doi.org/10.1103/PhysRevB.103.155103} {\bibfield  {journal} {\bibinfo
   {journal} {Phys. Rev. B}\ }\textbf {\bibinfo {volume} {103}},\ \bibinfo
  {pages} {155103} (\bibinfo {year} {2021}{\natexlab{b}})}\BibitemShut
  {NoStop}%
\bibitem [{\citenamefont {Balram}\ \emph {et~al.}(2022)\citenamefont {Balram},
  \citenamefont {Liu}, \citenamefont {Gromov},\ and\ \citenamefont
  {Papi\ifmmode~\acute{c}\else \'{c}\fi{}}}]{Balram21d}%
  \BibitemOpen
  \bibfield  {author} {\bibinfo {author} {\bibfnamefont {A.~C.}\ \bibnamefont
  {Balram}}, \bibinfo {author} {\bibfnamefont {Z.}~\bibnamefont {Liu}},
  \bibinfo {author} {\bibfnamefont {A.}~\bibnamefont {Gromov}},\ and\ \bibinfo
  {author} {\bibfnamefont {Z.}~\bibnamefont {Papi\ifmmode~\acute{c}\else
  \'{c}\fi{}}},\ }\bibfield  {title} {\bibinfo {title} {Very-high-energy
  collective states of partons in fractional quantum {Hall} liquids},\ }\href
  {https://doi.org/10.1103/PhysRevX.12.021008} {\bibfield  {journal} {\bibinfo
  {journal} {Phys. Rev. X}\ }\textbf {\bibinfo {volume} {12}},\ \bibinfo
  {pages} {021008} (\bibinfo {year} {2022})}\BibitemShut {NoStop}%
\bibitem [{\citenamefont {Dora}\ and\ \citenamefont {Balram}(2022)}]{Dora22}%
  \BibitemOpen
  \bibfield  {author} {\bibinfo {author} {\bibfnamefont {R.~K.}\ \bibnamefont
  {Dora}}\ and\ \bibinfo {author} {\bibfnamefont {A.~C.}\ \bibnamefont
  {Balram}},\ }\bibfield  {title} {\bibinfo {title} {Nature of the anomalous
  $4/13$ fractional quantum {Hall} effect in graphene},\ }\href
  {https://doi.org/10.1103/PhysRevB.105.L241403} {\bibfield  {journal}
  {\bibinfo  {journal} {Phys. Rev. B}\ }\textbf {\bibinfo {volume} {105}},\
  \bibinfo {pages} {L241403} (\bibinfo {year} {2022})}\BibitemShut {NoStop}%
\bibitem [{\citenamefont {Sharma}\ \emph {et~al.}(2023)\citenamefont {Sharma},
  \citenamefont {Pu}, \citenamefont {Balram},\ and\ \citenamefont
  {Jain}}]{Sharma22}%
  \BibitemOpen
  \bibfield  {author} {\bibinfo {author} {\bibfnamefont {A.}~\bibnamefont
  {Sharma}}, \bibinfo {author} {\bibfnamefont {S.}~\bibnamefont {Pu}}, \bibinfo
  {author} {\bibfnamefont {A.~C.}\ \bibnamefont {Balram}},\ and\ \bibinfo
  {author} {\bibfnamefont {J.~K.}\ \bibnamefont {Jain}},\ }\bibfield  {title}
  {\bibinfo {title} {Fractional quantum {Hall} effect with unconventional
  pairing in monolayer graphene},\ }\href
  {https://doi.org/10.1103/PhysRevLett.130.126201} {\bibfield  {journal}
  {\bibinfo  {journal} {Phys. Rev. Lett.}\ }\textbf {\bibinfo {volume} {130}},\
  \bibinfo {pages} {126201} (\bibinfo {year} {2023})}\BibitemShut {NoStop}%
\bibitem [{\citenamefont {Bose}\ and\ \citenamefont {Balram}(2023)}]{Bose23}%
  \BibitemOpen
  \bibfield  {author} {\bibinfo {author} {\bibfnamefont {K.}~\bibnamefont
  {Bose}}\ and\ \bibinfo {author} {\bibfnamefont {A.~C.}\ \bibnamefont
  {Balram}},\ }\bibfield  {title} {\bibinfo {title} {Prediction of non-abelian
  fractional quantum {Hall} effect at $\ensuremath{\nu}=2+\frac{4}{11}$},\
  }\href {https://doi.org/10.1103/PhysRevB.107.235111} {\bibfield  {journal}
  {\bibinfo  {journal} {Phys. Rev. B}\ }\textbf {\bibinfo {volume} {107}},\
  \bibinfo {pages} {235111} (\bibinfo {year} {2023})}\BibitemShut {NoStop}%
\bibitem [{\citenamefont {Sharma}\ \emph {et~al.}(2024)\citenamefont {Sharma},
  \citenamefont {Balram},\ and\ \citenamefont {Jain}}]{Sharma23}%
  \BibitemOpen
  \bibfield  {author} {\bibinfo {author} {\bibfnamefont {A.}~\bibnamefont
  {Sharma}}, \bibinfo {author} {\bibfnamefont {A.~C.}\ \bibnamefont {Balram}},\
  and\ \bibinfo {author} {\bibfnamefont {J.~K.}\ \bibnamefont {Jain}},\
  }\bibfield  {title} {\bibinfo {title} {Composite-fermion pairing at
  half-filled and quarter-filled lowest {Landau} level},\ }\href
  {https://doi.org/10.1103/PhysRevB.109.035306} {\bibfield  {journal} {\bibinfo
   {journal} {Phys. Rev. B}\ }\textbf {\bibinfo {volume} {109}},\ \bibinfo
  {pages} {035306} (\bibinfo {year} {2024})}\BibitemShut {NoStop}%
\bibitem [{\citenamefont {Balram}\ and\ \citenamefont
  {Regnault}(2024)}]{Balram24}%
  \BibitemOpen
  \bibfield  {author} {\bibinfo {author} {\bibfnamefont {A.~C.}\ \bibnamefont
  {Balram}}\ and\ \bibinfo {author} {\bibfnamefont {N.}~\bibnamefont
  {Regnault}},\ }\href@noop {} {\bibinfo {title} {Fractional quantum {Hall}
  effect of partons and the nature of the 8/17 state in the zeroth {Landau}
  level of bilayer graphene}} (\bibinfo {year} {2024}),\ \Eprint
  {https://arxiv.org/abs/2404.15547} {arXiv:2404.15547 [cond-mat.str-el]}
  \BibitemShut {NoStop}%
\bibitem [{\citenamefont {Wen}(1991)}]{Wen91}%
  \BibitemOpen
  \bibfield  {author} {\bibinfo {author} {\bibfnamefont {X.~G.}\ \bibnamefont
  {Wen}},\ }\bibfield  {title} {\bibinfo {title} {Non-abelian statistics in the
  fractional quantum {Hall} states},\ }\href
  {https://doi.org/10.1103/PhysRevLett.66.802} {\bibfield  {journal} {\bibinfo
  {journal} {Phys. Rev. Lett.}\ }\textbf {\bibinfo {volume} {66}},\ \bibinfo
  {pages} {802} (\bibinfo {year} {1991})}\BibitemShut {NoStop}%
\bibitem [{\citenamefont {Umucal\ifmmode \imath \else~\i \fi{}lar}\ \emph
  {et~al.}(2018)\citenamefont {Umucal\ifmmode \imath \else~\i \fi{}lar},
  \citenamefont {Macaluso}, \citenamefont {Comparin},\ and\ \citenamefont
  {Carusotto}}]{Umucalilar18}%
  \BibitemOpen
  \bibfield  {author} {\bibinfo {author} {\bibfnamefont {R.~O.}\ \bibnamefont
  {Umucal\ifmmode \imath \else~\i \fi{}lar}}, \bibinfo {author} {\bibfnamefont
  {E.}~\bibnamefont {Macaluso}}, \bibinfo {author} {\bibfnamefont
  {T.}~\bibnamefont {Comparin}},\ and\ \bibinfo {author} {\bibfnamefont
  {I.}~\bibnamefont {Carusotto}},\ }\bibfield  {title} {\bibinfo {title}
  {Time-of-flight measurements as a possible method to observe anyonic
  statistics},\ }\href {https://doi.org/10.1103/PhysRevLett.120.230403}
  {\bibfield  {journal} {\bibinfo  {journal} {Phys. Rev. Lett.}\ }\textbf
  {\bibinfo {volume} {120}},\ \bibinfo {pages} {230403} (\bibinfo {year}
  {2018})}\BibitemShut {NoStop}%
\bibitem [{\citenamefont {Macaluso}\ \emph {et~al.}(2020)\citenamefont
  {Macaluso}, \citenamefont {Comparin}, \citenamefont {Umucal\ifmmode \imath
  \else~\i \fi{}lar}, \citenamefont {Gerster}, \citenamefont {Montangero},
  \citenamefont {Rizzi},\ and\ \citenamefont {Carusotto}}]{Macaluso20}%
  \BibitemOpen
  \bibfield  {author} {\bibinfo {author} {\bibfnamefont {E.}~\bibnamefont
  {Macaluso}}, \bibinfo {author} {\bibfnamefont {T.}~\bibnamefont {Comparin}},
  \bibinfo {author} {\bibfnamefont {R.~O.}\ \bibnamefont {Umucal\ifmmode \imath
  \else~\i \fi{}lar}}, \bibinfo {author} {\bibfnamefont {M.}~\bibnamefont
  {Gerster}}, \bibinfo {author} {\bibfnamefont {S.}~\bibnamefont {Montangero}},
  \bibinfo {author} {\bibfnamefont {M.}~\bibnamefont {Rizzi}},\ and\ \bibinfo
  {author} {\bibfnamefont {I.}~\bibnamefont {Carusotto}},\ }\bibfield  {title}
  {\bibinfo {title} {Charge and statistics of lattice quasiholes from density
  measurements: A tree tensor network study},\ }\href
  {https://doi.org/10.1103/PhysRevResearch.2.013145} {\bibfield  {journal}
  {\bibinfo  {journal} {Phys. Rev. Res.}\ }\textbf {\bibinfo {volume} {2}},\
  \bibinfo {pages} {013145} (\bibinfo {year} {2020})}\BibitemShut {NoStop}%
\end{thebibliography}%
\end{document}